\documentclass[11pt]{article}
\usepackage{amsmath}
\usepackage{times}
\usepackage{amsfonts,amssymb,float, mathtools,multirow, amsthm}
\usepackage{amsmath}

\usepackage{bm,xcolor}
\newcommand{\bs}[1] {\bm{#1}}
\usepackage{natbib}
\usepackage{xr}
\usepackage{enumitem}
\usepackage{graphicx,url}
\usepackage[noblocks]{authblk}

\DeclareMathOperator*{\E}{\rm E}
\usepackage[plain,noend]{algorithm2e}

\makeatletter
\renewcommand{\algocf@captiontext}[2]{#1\algocf@typo. \AlCapFnt{}#2} 
\def\@algocf@capt@plain{top}
\renewcommand{\algocf@makecaption}[2]{%
  \addtolength{\hsize}{\algomargin}%
  \sbox\@tempboxa{\algocf@captiontext{#1}{#2}}%
  \ifdim\wd\@tempboxa >\hsize
    \hskip .5\algomargin%
    \parbox[t]{\hsize}{\algocf@captiontext{#1}{#2}}
  \else%
    \global\@minipagefalse%
    \hbox to\hsize{\box\@tempboxa}
  \fi%
  \addtolength{\hsize}{-\algomargin}%
}
\makeatother

\newtheorem{theorem}{Theorem}[section]

\def\E{\mathbb{E}}

\begin{document}




\title{ Robust Deep Neural Network Estimation for Multi-dimensional Functional Data}

\author[1]{Shuoyang Wang}
\author[1]{Guanqun Cao}
\author[2]{for the Alzheimer's
Disease Neuroimaging Initiative}

\affil[1]{Department Mathematics and Statistics, Auburn University, U.S.A.}

\date{}

\maketitle

\begin{quotation}
\noindent \textit{Abstract:} In this paper, we propose a   robust estimator for the location function from multi-dimensional functional data. The proposed estimators are based on the deep neural networks with ReLU activation function. At the meanwhile, the estimators are less susceptible to outlying observations and model-misspecification. For any multi-dimensional functional data, we provide the uniform convergence rates for the proposed robust deep neural networks estimators.  Simulation studies   illustrate the competitive performance of the robust deep neural network  estimators on regular data and their superior performance on data that contain anomalies.
The proposed method is also applied to
analyze 2D and 3D images of patients with Alzheimer's disease
obtained from the Alzheimer Disease Neuroimaging Initiative database.

\vspace{9pt} \noindent \textit{Key words and phrases:}  
 {Functional data analysis};
 {Functional neural networks};
{M-estimators};
{Rate of convergence};
{ReLU activation function};
 {ADNI databas}.
\end{quotation}

%

\pagestyle{myheadings}
\thispagestyle{plain}
{}


\section{Introduction}
\label{s:intro}

We consider the problem of robust estimation of the location function   from a collection of functional observations defined over $\mathbb{R}^d$ ($d\geq 1$)
a multi-dimensional domain. To be precise, let $\xi = \{\xi(\mathbf{X}) : \mathbf{X} \in \mathcal{I}\}$ be a compactly supported random
field, i.e., a real-valued second-order stochastic process on a compact set $\mathcal{I}\subset \mathbb{R}^d$. Such data are nowadays commonly
referred to as functional data.
In many applications, data are collected over one-dimensional domains (i.e., $d = 1$) such as  time-varying trajectories   and relevant research has been enjoying considerable popularity.  The readers are referred to some monographs  \cite{Ramsay:Silverman:05,Wang:etal:16,Kokoszka:Reimherr:17,HsingEubank2015} for a comprehensive overview of functional data analysis (FDA).  Thanks to the improved capabilities of data
recording and storage, as well as advances in scientific computing and data science, many new forms of functional data have emerged. Instead of traditional unidimensional functional data, multi-dimensional functional data becomes increasingly common in various fields, such as geographical science and neuroscience. For example,  for the early detection and tracking of Alzermer's diease, the Alzheimer's Disease Neuroimaging Initiative (ADNI) database (\url{adni.loni.usc.edu}) contains  each individual's 3D brain-scans.   Despite the promising of multi-dimensional functional data, statistical methods for such data are limited, except for very few existing
works, for example, \cite{Chen:Muller:12,Zhou:Pan:14,Chen:Jiang:17}.

A fundamental problem in FDA is the estimation of central tendency,  yet most current estimation procedures either  lack robustness with respect to the many kinds of anomalies one can encounter in the functional setting or only focus on the robustness for unidimensional scenario.  The fact that robust estimation has not been widely investigated for multi-dimensional scenario is certainly not owing to lack of interesting applications, but to the greater technical difficulty to handle such loss function for multi-dimensional functional data and establish their theoretical properties.

To give some background on our proposed method for multi-dimensional functional
data, we first review several relevant robust FDA methods that have been developed for analyzing unidimensional functional
data.  \cite{bali2011robust,lee2013m} proposed robust estimators for the functional principal components by adapting
the projection pursuit approach and based on MM estimation, respectively. \cite{maronna2013robust} established a robust version
of spline-based estimators for a linear functional regression model. \cite{shin2015rkhs} proposed a robust procedure based on convex and non-convex loss functions in
functional linear regression models. Recently, \cite{Lima:etal:19,Lima:etal:19b}  proposed  robust estimators and  associated simultaneous confidence
bands for the mean function of functional data using least absolute deviation and M-estimation, respectively.

We notice that there are few exiting works on  robust  methods for analyzing  so-called two-way functional data which consist of a data matrix whose row and column
domains are both structured,  as when the
data are time series collected at different locations in space. For example, \cite{Zhang:etal:AOAS:13}  develop a robust regularized singular value decomposition
  method for analyzing such special type functional data.  It is formulated as
a penalized loss minimization problem and a pre-decided  two-way roughness penalty function is used to ensure smoothness along each of the two functional domains. As this method is only designed for the special two-way functional data, it can not be adopted to the  general multi-dimensional FDA directly. Furthermore,  a lack of theoretical analysis provides inadequate assurance to robust methods practitioners.

To remedy these deficiencies, we introduce the first class of optimal robust location estimators based on the deep neural network  (DNN) method.
DNN is one of the most promising and vibrate areas in deep learning. 
DNN has been recently applied in various nonparametric regression problems recently, they have been shown to successfully overcome the curse of dimensionality in
nonparametric regression; see \cite{Schmidt:19,Bauer:Kohler:19, Liu:19,Liu:etal:2021}. There are also some   works proposed for deep learning algorithms for FDA from the statistical point of view \citep{Thind:etal:20a}.
Based on the sparsely connected DNN,  \cite{wang2021stat} proposed a DNN estimator for the mean function from functional data  based on  the least squares neural network regression.  However, none of them works on the robust statistics, not to mention the proven theoretical results for robust FDA.


 The contributions of the present paper are three-fold. First, to the best of our knowledge, this is the first work on proposing DNN based robust estimator for FDA.  We propose a broad class of M-type  RDNN (robust DNN) estimators to estimate location functions for multi-dimensional functional data.
 Second, RDNN estimators come with theoretical guarantees. In particular, we   study the rate of convergence of the estimator under weak assumptions  and show that the estimator is  rate-optimal even for any $d$-dimensional functional data. By borrowing the strength from the DNN, the convergence rate of the  proposed RDNN estimator does not depend on the dimension $d$. Finally,   our analyses are fully nonparametric. At the meanwhile, RDNN estimator does not suffer  the curse-of-dimensionality which is a classical drawback in the traditional nonparametric regression framework.

The paper is structured as follows.
 Section \ref{SEC:FDA} provides the model setting in FDA and  introduces multilayer feed-forward
artificial neural networks and discusses mathematical modeling. The implementation on hyperparameter selections also be included in  Section \ref{SEC:FDA}. The theoretical properties of the proposed RDNN estimator can
be found in Section \ref{SEC:main}. Section \ref{SEC:Implementation} provides the detailed implementation on neural network's architecture selecting and  training.  In Section \ref{SEC:sim}, it is shown that the finite sample performance of  proposed neural network estimator. The proposed method is applied to the spatially normalized positron emission tomography (PET) data from ADNI in Section \ref{SEC:realdata} and make some concluding remarks in Section \ref{SEC: discussion}. Technical proofs are collected in the Appendix.

\section{The model  and the robust deep neural network  estimator}
\label{SEC:FDA}


\subsection{FDA model}
Let us first assume the process $\left\{ \xi (\mathbf{X}), \mathbf{X}\in \left[ 0,1\right]^d
\right\} $  is $L^{2}$, i.e., $E\int_{\left[ 0,1\right] ^d }\xi
^{2}(\mathbf{X})d\mathbf{X}<+\infty $. In the classical FDA setting, $d=1$ refers to the index variable as time. When $d=2, 3$, it could also be a spatial variable, such as  in image or geoscience applications.
We model  the multi-dimensional functional data as noisy sampled points from a collection of trajectories that are assumed to be independent realizations of a smooth random function $\xi (\mathbf{X})$, with unknown mean function $f_0(\mathbf{X})=\E\{\xi\left(\mathbf{X}\right) \}$.
We consider a  version of the model that incorporates uncorrelated measurement
errors.
Let   $\xi
_{1},\ldots,\xi_n$ denote $n$  independent and identically distributed (i.i.d.) copies of $\xi$ at  points $\mathbf{X}=(X_{1}, \ldots, X_{d})$, $1\leq i\leq n$.  Our goal is to   recover the mean function $f_0(\mathbf{X}_j)$ from the noisy observations of the discretized functional data:
\begin{eqnarray} \label{DEF:model}
Y_{ij} =\xi_i\left(\mathbf{X}_j\right) + e_{i}\left(\mathbf{X}_j\right), ~~i = 1, 2, \ldots, n, j = 1, 2, \ldots, N_i, 
\end{eqnarray}
where $e_{i}\left(\mathbf{X}_j\right)$ are random noise variables. In \cite{Yao:etal:05b,Cao:Yang:Todem:12,Cao:Wang:Li:Yang:14}, it is assumed that the noise variables $e_{i}\left(\mathbf{X}_j\right)$ are independent of the $\xi_i$ and i.i.d.  with zero mean and finite
variance. However, we allow for correlated errors that are not necessarily independent of
the functional curves.

In terms of mean-deviations, model (\ref{DEF:model}) can be equivalently written as
\begin{eqnarray} \label{DEF:model2}
Y_{ij}  =f_0\left(\mathbf{X}_j\right) +   \epsilon_{i}\left(\mathbf{X}_j\right), ~~i = 1, 2, \ldots, n, j = 1, 2, \ldots, N_i,
\end{eqnarray}
where  $\epsilon_i(\mathbf{X}_j)=\xi_i\left(\mathbf{X}_j\right)-\E\{\xi_i\left(\mathbf{X}_j\right)\}+e_i(\mathbf{X}_j)$
 denotes the error process associated
with the $i$-th response evaluated at $\mathbf{X}_{j}$. The problem is thus reformulated as a regression
problem with repeated measurements and possibly correlated errors. In the following, for simple notations, we consider the equally number of observations for each subject design ($N_i  \equiv N$). The main results can be easily extended to irregular number design. 
\subsection{Robust deep neural network  estimator}
We first briefly introduce the necessary  notations   and terminologies used in the neural networks.    Popular choice of activation functions includes rectified linear unit (ReLU), sigmoid, and tanh. In this article, we will mainly focus on neural networks with the ReLU activation function, i.e.,  $\sigma(x)=(x)_+$ for $x\in\mathbb{R}$.
For any real vector  $\bs{y}=(y_1,\ldots,y_r)^\top$, define the shift activation function $\sigma (\bs{y})=(\sigma(y_1),\ldots,\sigma(y_r))^\top$.
For an integer $L\ge1$ and $\bs{p}=(p_0,p_1,\ldots,p_{L}, p_{L+1})\in\mathbb{N}^{L+2}$, let $\mathcal{F}(L, \bs{p})$ denote the class of DNN, with $L$ hidden layers and $p_l$ nodes on hidden layer $l$, for $l=1,\ldots,L$. 
We consider the feed-forward neural network class, and any $f\in\mathcal{F}(L, \bs{p})$ has a composition structure, i.e.,
 \begin{equation} \label{EQ:f}
f(\mathbf{x}) = \mathbf{W}_L \sigma\left( \mathbf{W}_{L-1}\ldots \sigma \left( \mathbf{W}_1 \sigma  \left(\mathbf{W}_0\mathbf{x}+\mathbf{u}_0\right)+\mathbf{u}_1\right)+\ldots+\mathbf{u}_{L-1}\right)+\mathbf{u}_{L} , \,\,\,\,\mathbf{x}\in\mathbb{R}^d,
\end{equation}
where $\mathbf{W}_l\in\mathbb{R}^{p_{l+1}\times p_{l}}$ are weight matrices and $\bs{u}_l\in\mathbb{R}^{p_l}$
are shift vectors, for $l=1,\ldots,L$.  Owing to the large capacity of neural network class, it tends to overfit the training dataset easily. To avoid the overfitting and reduce the computational burden, we train the robust estimator using the following  s-sparse ReLU DNN  class:
\begin{eqnarray}\label{EQ:class}
&&\hspace{3mm} \mathcal{F}(  L, \bs{p}, s ) \nonumber \\
 &=& \left\{ f\in \mathcal{F}( L, \bm{p}) : \sum_{l=0}^L\| \mathbf{W}_l\|_0 + \|\mathbf{u}_l\|_0 \leq s,  \max_{l=0,\ldots,L}\| \mathbf{W}_l\|_{\infty} + \|\mathbf{u}_l\|_{\infty} \leq 1, \right. \nonumber \\
 &&\left. \hspace{2.2cm}\|f\|_\infty\le 1 \right\},
\end{eqnarray}
where  $ \| \cdot\|_{\infty}$ denotes the maximum-entry norm of a matrix/vector
or supnorm of a function, $\| \cdot \|_0$  denotes the number of non-zero entries of a matrix or vector, $s>0$ controls the number of nonzero weights and shift.
The selecting procedures of  unknown tuning parameters $(L, \mathbf{p}, s)$ shall be given in  Section \ref{SEC:Implementation}. To simplify the notations, we write $\mathcal{F}$ instead of $\mathcal{F}(  L, \bs{p}, s )$ in the following.

In the  regression model,  the common objective is to find an optimal estimator by minimizing a loss function. In the DNN setting, this coincides with training neural networks by minimizing the empirical risk  over all the training data. In particularly, given the networks in (\ref{EQ:class}), the proposed  RDNN   estimator  is defined as
\begin{equation}\label{EQ:fhat}
\widehat{f} = \arg\min_{f\in  \mathcal{F} }  \frac{1}{nN}\sum_{i = 1}^n\sum_{j = 1}^N\rho \left( {Y}_{i j}- f\left(\mathbf{X}_j\right)\right) ,
\end{equation}
where $\rho$ is some convex nonnegative loss function  satisfying $\rho(0) = 0$ and $\mathcal{F}$ is some function class.  This formulation is very general, allowing the flexibility in the choice of the
loss function, so that better resistance towards outlying observations is achieved.  One of the well-known examples of such loss functions is  Huber’s loss function given by $\rho_k(x) = x^2/2\mathbb{I}(|x|\leq k)+k(|x|-k/2)\mathbb{I}(|x|> k)$, where $\mathbb{I}(\cdot)$ is the indicator function, and $k > 0$ controls the blending of square and absolute losses.    
Furthermore,   the symmetry of the loss function in (\ref{EQ:fhat}) is not required, such versatile estimators may be readily incorporated into the present framework. Indeed, to estimate conditional quantiles,   one would only need to select the loss function as $\rho (x)=x(\tau-\mathbb{I}(x<0))$ for some $\tau\in(0,1)$. Finally, the asymptotic properties of
quantile   estimators are covered by the theory developed
in Section \ref{SEC:main}.


\section{Theoretical properties of the RDNN estimator}
\label{SEC:main}
\subsection{Definitions and notations}
Define the ball of $\beta$-H\"{o}lder functions with radius $K$ as
\begin{eqnarray*}
\mathcal{C}_d^{\beta}(D, K) = &&\left\{ \right.  f: D \subset \mathbb{R}^d \rightarrow \mathbb{R} : \\
&&\left. \sum_{\bs{\alpha}:|\bs{\alpha}|<\beta} \| \partial^{\bs{\alpha}}f\|_{\infty} + \sum_{\bs{\alpha}:|\bs{\alpha}|=\lfloor{\beta}\rfloor}\sup_{\bs{x}, \bs{y} \in D, \bs{x} \neq \bs{y} } \frac{|\partial^{\bs{\alpha}}f(\bs{x}) - \partial^{\bs{\alpha}}f(\bs{y}) |}{|\bs{x} - \bs{y}|_{\infty}^{\beta - \lfloor\beta\rfloor}} \leq K \right\},
\end{eqnarray*}
where $\partial^{\bs{\alpha}}$ = $\partial^{\alpha_1}\ldots\partial^{\alpha_d}$ with $\bs{\alpha}$ = $(\alpha_1, \ldots, \alpha_d) \in \mathbb{N}^d$ and $|\bs{\alpha}|:= |\bs{\alpha}|_1$.

We assume the true location function $f_0$ has  the natural composition structure, i.e.,
\begin{equation*}
f_0 = g_q \circ g_{q-1} \circ \ldots \circ g_1 \circ g_0,
\end{equation*}
where $g_\ell : \left[a_\ell, b_\ell \right]^{d_\ell} \rightarrow \left[a_{\ell+1}, b_{\ell+1} \right]^{d_{\ell+1}} $,   $g_\ell = \left( g_{\ell j}\right)^{\top}_{j=1, \ldots, d_{\ell+1}}$, $\ell=1,\ldots,q$, with unknown parameters $d_\ell$ and $q$. We assume each  $g_{\ell j}$  is $\beta_\ell$-H\"{o}lder function with radius $K_\ell$. Let $t_\ell$ be the maximal number of variables  on which each of the  $g_{\ell j}$ depends on $t_\ell$, and $t_\ell \leq d_\ell$. Since $g_{\ell j}$ is also $t_\ell$-variate,   the true underlying function space becomes
\begin{eqnarray}\label{EQ:gfunction}
&&\mathcal{G}\left(q, \mathbf{d}, \mathbf{t}, \bs{\beta}, \mathbf{K} \right)\nonumber \\
&:=& \left\{ \right. f = g_q \circ \ldots \circ g_0 :   g_\ell = (g_{\ell j})_j : \left[a_\ell, b_\ell \right]^{d_\ell} \rightarrow \left[a_{\ell+1}, b_{\ell+1} \right]^{d_{\ell+1}}, \nonumber \\
&& g_{\ell j} \in \mathcal{C}^{\beta_\ell}_{t_\ell}\left( \left[a_\ell, b_\ell\right]^{t_\ell}, K_\ell\right), |a_\ell|, |b_\ell| \leq K_\ell \left. \right\},
\end{eqnarray}
with $\mathbf{d} := (d_0, \ldots, d_{q+1})$,  $\mathbf{t} := (t_0, \ldots, t_q)$,  $\bs{\beta} := (\beta_0, \ldots, \beta_q)$, $\mathbf{K} := (K_0, \ldots, K_q)$ and
$
\beta_\ell^{\ast} := \beta_\ell \prod_{k=\ell+1}^q (\beta_k \wedge 1)
$.
\subsection{Assumptions}
In this section, we develop the convergence rate of the proposed RDNN estimator in (\ref{EQ:fhat}).
For simple notations, $\log$ denotes the
 logarithmic function with base $2$.
For  sequences $(a_n)_n$ and $(b_n)_n$,  $a_n \asymp b_n$ means $a_n \leq c_1 b_n$  and $a_n \geq c_2 b_n$ where $c_1$ and $c_2$ are  absolute constants for any $n$.


We now introduce the main assumptions:
\begin{description}
\item[(A1)]  The true regression function $f_0 \in \mathcal{G}\left(q, \mathbf{d}, \mathbf{t}, \bs{\beta}, \mathbf{K} \right)$.
\item[(A2)]  The RDNN estimator $\widehat{f} \in \mathcal{F}(L, \mathbf{p}, s)$, where  $L \asymp \log (nN^{\nu})$, $s \asymp (nN^{\nu})^{\frac{1}{\theta+1}}$,  $\min_{l=1, \ldots, L}p_l \asymp  (nN^{\nu})^{\frac{1}{\theta+1}}$, for
$\theta = \min_{\ell=0,\ldots,q}\frac{2\beta_\ell^{\ast}}{t_\ell}$ and $\nu\geq 0$.
\item[(A3)] The loss function $\rho(\cdot)$ is an absolutely continuous convex function on $\mathbb{R}$ with derivative $\psi(\cdot):=\rho^{\prime}(\cdot)$ existing almost everywhere.
\item[(A4)] There exist finite constants $\kappa$ and $c_1$ such that for all $x\in\mathbb{R}$ and $|x^{\prime}|<\kappa$, $|\psi(x+x^{\prime})-\psi(x )|<c_1$.
\item[(A5)] There exist a finite constant $c_2$ such that $\sup_{j\leq N}\E\{|\psi(\epsilon_{1j}+u)-\psi(\epsilon_{1j}) |^2\}<c_2|u|$, as $|u|\rightarrow 0$.
\item[(A6)]   $\sup_{j\leq N}\E\{(\psi(\epsilon_{1j})) ^2 \}=O(N^{-\nu})$, for some constant $\nu\geq 0$, and $ \E\{\psi(\epsilon_{1j})  \}=0$. There exist finite constants $\delta_j$, $j=1,\ldots,N$ such that $0<\inf_{j \leq N}\delta_j\leq\sup_{j \leq N}\delta_j<\infty$ and $\sup_{j\leq N}|\E\{ \psi(\epsilon_{1j}+u)\}-\delta_j u|=o(u)$, as $|u|\rightarrow 0$.

\end{description}

Assumption (A1)  is a natural definition for the neural network, which is fairly flexible and many well known function classes are contained in it.  For example, the generalized additive model
   $f_0(\mathbf{x}) = h\left(\sum_{i=1}^df_i(x_i)\right)$,
  can be written as a composition of three functions
 $f_0 = g_2\circ g_1 \circ g_0$, with $g_0(x_1,\ldots,x_d)=(f_1(x_1),\ldots,f_d(x_d))$, $g_1(x_1,\ldots,x_d)=\sum_{i=1}^dx_i$, and $g_2(\cdot) = h(\cdot)$.
Assumption (A2) depicts the architecture and parameters' setting  in the network space.
To use discontinuous score functions, Assumptions (A3)-(A6)  impose some regularity on the error process and its finite-dimensional distributions.  In particularly, Assumptions (A3) guarantees the existence of the solution
of the optimization problem in (\ref{EQ:fhat}). Most of the  loss
functions chosen in practice satisfy this condition, such as the Huber loss function.  Assumptions (A4) and (A5) require  boundedness
and some regularity of the score function, which    are   standard conditions for M-estimation procedures for FDA, see the similar conditions required in \cite{Lima:etal:19}. For the first part of Assumption (A6), when considering the classical $L_2$ loss, it essentially makes sure the maximal value of the covariance function is finite and decreases  when the number of measurements increases. They are standard regularity conditions for the covariance functions in FDA  literature, see \cite{Cao:Wang:Li:Yang:14, Lima:etal:19,wang2021stat} for example.
The second part of Assumption
(A6) essentially requires that for any $j=1,\ldots,N$,   function $h_j(u)= E\{\psi(\epsilon_{1j}+u)\}$,
is differentiable with strictly positive derivative at the origin. This is a necessary condition
for the minimum to be well-separated in the limit. Assumption
 (A6) on the score function $\psi$ is also standard conditions in   M-estimation for functional data literature, see \cite{Lima:etal:19,Kalogrids:20}. It is also not stringent assumptions for errors, for example, $\epsilon_{ij}$'s following a zero mean Gaussian process or
mixture Normal–Cauchy distribution. We provide more detailed examples for $\epsilon_{ij}$'s  in Section \ref{SEC:sim}.


\subsection{Unified rate of convergence}

The following theorem establishes the unified convergence rate of the RDNN estimator $\widehat{f}$ for any multi-dimensional functional data under the empirical norm. Its proof and some technical lemmas are provided in the Appendix.

\begin{theorem} \label{THM: rate}
Under Assumptions (A1)-(A6), we have
\begin{equation}\label{EQ: convergence rate}
\|\widehat{f} - f_{0} \|_N^2 = O_p(nN^{\nu})^{-\frac{\theta}{\theta+1}}\log^6 (nN^{\nu}),
\end{equation}
where  ${\nu} \geq 0$, $\theta = \min_{\ell=0,\ldots,q}\frac{2\beta_\ell^{\ast}}{t_\ell}$. 
\end{theorem}
It is interesting to observe that our proposed RDNN estimator enjoys the same asymptotic  as the least squares  DNN estimator \cite{wang2021stat} does. Specifically,
the convergence rate for M-type  DNN estimator in the functional  regression model depends on the smoothness, i.e., $\beta^*_{\ell}$,  and the intrinsic dimension, i.e., $t_{\ell}$, of the true mean function $f_0$, and the decay rate of the maximal value of the  variance function $\E\{(\psi(\epsilon_{1j})) ^2 \}$.

%

\section{Implementation}
\label{SEC:Implementation}
Different from classical nonparametric estimators, $\widehat{f}$ has no analytical expression or basis expansion expression.
The proposed robust estimator is constructed using the neural network class which is fully characterized by the architectures $(L, \bs{p}, s)$. We now provide the  detailed implementation procedure for the proposed  estimator in (\ref{EQ:fhat}).

\subsection{Neural network's architecture selection}\label{SEC:architecture}
 In the DNNs' computations,  tuning parameters are crucial as they control the overall behavior of the proposed estimator and the learning process. The tuning parameters are so-called network architecture parameters, which include the number of layers  $L$, the number of hidden neurons within these layers  $\bs{p}$, and sparse parameter $s$. There are fairly rich literature discussing the optimization selection, such as grid search, random search, and Bayesian optimization. Nevertheless, the selection of network architecture parameters has been rarely discussed. In practice, some model selection methods such as cross-validation may have good performances, but with huge computational burdens.
    For this reason, considering both the computational efficiency and the theoretical guarantee, we select architecture parameters  based on the assumptions in Theorem \ref{THM: rate}. Particularly,  we  choose $L =\lceil 0.5\log (nN^{1/2})\rceil$, $ p_l = \lceil 10n^{1/2}N^{1/4}\rceil$, $s = \lceil 5n^{1/2}N^{1/4}\rceil L$. We also select   $\nu=\theta=0.5$ in Assumption (A2). This   choice of $\nu$ and $\theta$ includes a large scope of true function classes. Note that in our considered sparse neural network space $\mathcal{F}$, the sparse parameter $s$ should be carefully selected. When designing the network architecture practically, the dropout rate is suggested as $  \lceil 5n^{1/2}N^{1/4}\rceil( \lceil 10n^{1/2}N^{1/4}\rceil)^{-1}$ in each layer during the optimization procedure.  

 \subsection{Training neural networks}\label{SEC:training}
 The minimization in (\ref{EQ:fhat}) is generally a computational cumbersome optimization
problem owing to non-linearities and non-convexities. The most commonly used solution
utilises stochastic gradient descent (SGD) to train a neural network. SGD uses a batch of
a specific size, that is, a small subset of the data (typical size $B=2^2$ to $2^{10}$) is randomly drawn at each iteration of optimization to evaluate the gradient, to alleviate the computation hurdle.  Our input size of network is $nN$, thus we choose relatively large batches $B$ from $256$ to $512$ depending on the performance of convergence.  A pass of the whole training set is called an epoch. Typical choices of epochs are $200$, $300$ and $500$. The number of epochs
defines the number of times that the learning algorithm works through the entire training dataset. The step of the derivative at each epoch is controlled by the learning rate which is $0.001$. The readers are referred to recent monographs (\cite{Fan:etal:19}) for a general discussion of these numerical challenges. There are certainly some challenges for SGD to train DNN. For example, albeit good theoretical guarantees for well-behaved problems, SGD
might converge very slowly; the learning rates are difficult to tune (\cite{Zhu:etal:19}). To overcome these challenges, we use a variant gradient based optimization algorithm Adam. Different from the classical SGD procedures, Adam is a method for efficient stochastic optimization that only requires first-order gradients with little memory requirement. Hence, it is well suited for problems when there are large sample sizes and parameters (\cite{Kingma:Ba:15}), and is widely used in network training for FDA, such as \cite{wang2021stat}. In our numerical studies, Adam provides the best results and is the most computationally efficient among other gradient based algorithms. In the real applications, we recommend Adam algorithm for finding RDNN estimators in (\ref{EQ:fhat}).

\section{Simulation}
\label{SEC:sim}

To illustrate the finite sample performance of the introduced RDNN estimators based on our proposed
neural networks method, we  conduct substantial simulations for both 2D  and 3D  functional data.  All experiments are conducted in \texttt{R}.   We summarize  \texttt{R} codes and examples for the proposed RDNN algorithms   on \texttt{GitHub}  (\url{https://github.com/FDASTATAUBURN/RDNN}).

 \subsection{2D simulation}\label{subSEC:2D}
  The 2D functional data are generated from the model:
\begin{equation} \label{EQ:sim}
Y_{ij} =f_0\left(\mathbf{X}_j\right) +  \epsilon_{ij},
\end{equation}
where the true mean function $f_0(\mathbf{x}_{j})=-8\left[1+\exp\left\{ \cot(x_{1j}^2)\cos(2\pi x_{2j})\right\}\right]^{-1}$, and  $\mathbf{x}_j=  \left(j_1/N_2, j_2/ N_2 \right)$, $1 \leq j_1, j_2 \leq N_2$, are the equally spaced grid points on $\left[ 0, 1\right] ^2$, and $N_2^2=N$. The error term is $ \epsilon_{ij}=\eta(\mathbf{X}_j)+e_{ij}$, where $\eta(\cdot)$ is generated from a Guassian process, with zero mean and covariance function $G_0(\mathbf{x}_j,\mathbf{x}_{j'})=\sum_{k=1}^2\cos(2\pi(x_{kj}-x_{kj'}))$, $j,j'=1,\ldots,N$. The measurement errors
$e_{ij}$'s are i.i.d. standard normal random variables.

\begin{figure}
\begin{center}
\hspace{-9.2cm}  $f_0$ \hspace{0.25cm}
$\begin{array}{l}
\includegraphics[width = 0.2\textwidth]{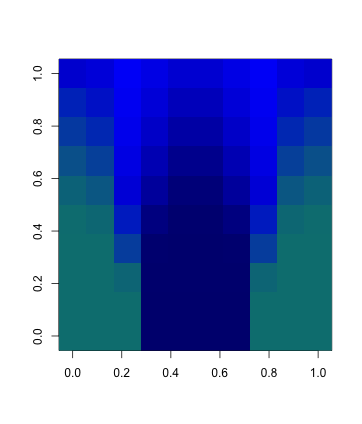}
\includegraphics[width=0.047 \textwidth]{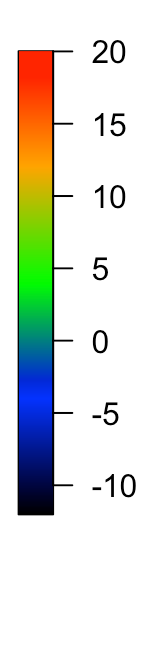}
\end{array}$\\
outliers
$\begin{array}{l}
\includegraphics[width = 0.2\textwidth]{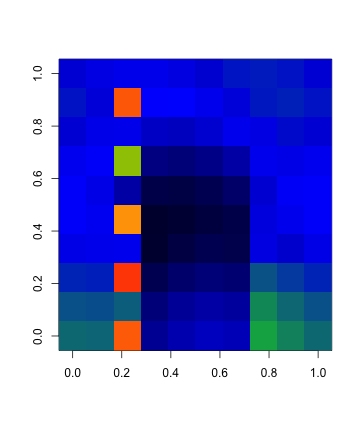}
\includegraphics[width = 0.2\textwidth]{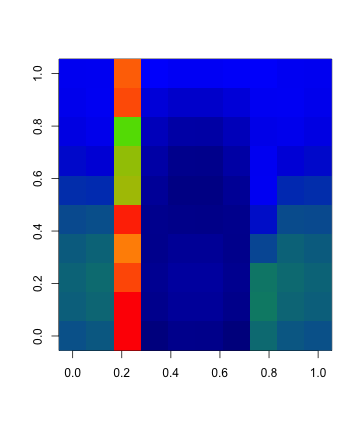}
\includegraphics[width = 0.2\textwidth]{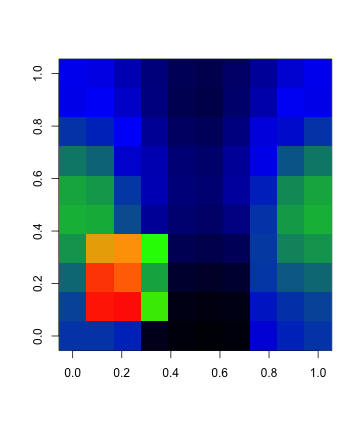}
\includegraphics[width = 0.2\textwidth]{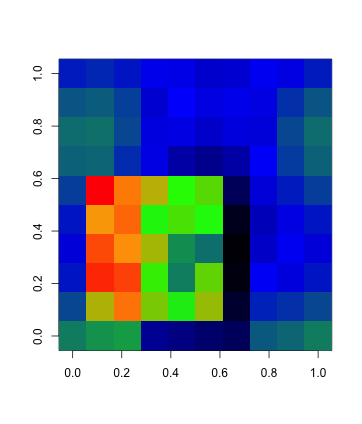}
\includegraphics[width=0.047 \textwidth]{colorbar_2.png}
\end{array}$\\
DNN \;\;
$\begin{array}{l}
\includegraphics[width = 0.2\textwidth]{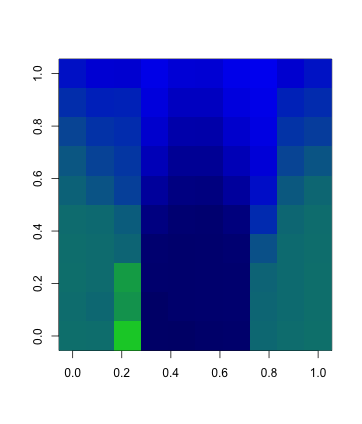}
\includegraphics[width = 0.2\textwidth]{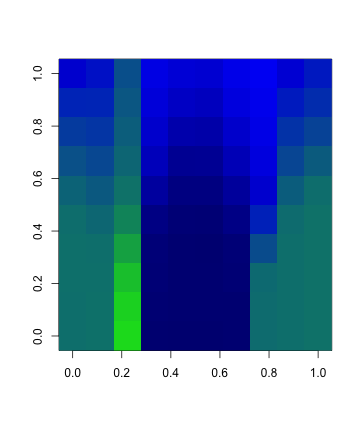}
\includegraphics[width = 0.2\textwidth]{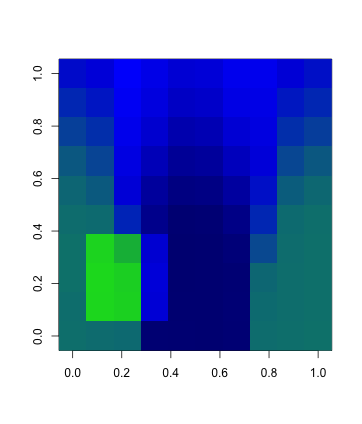}
\includegraphics[width = 0.2\textwidth]{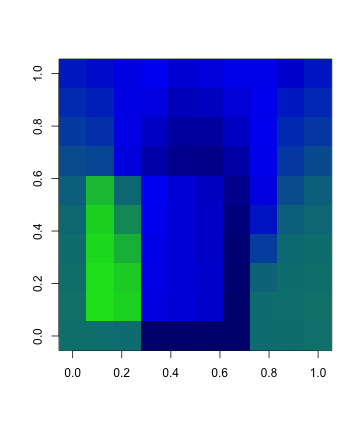}
\includegraphics[width=0.047 \textwidth]{colorbar_2.png}
\end{array}$\\
RDNN
$\begin{array}{l}
\includegraphics[width = 0.2\textwidth]{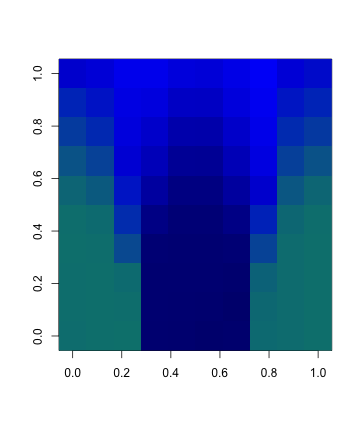}
\includegraphics[width = 0.2\textwidth]{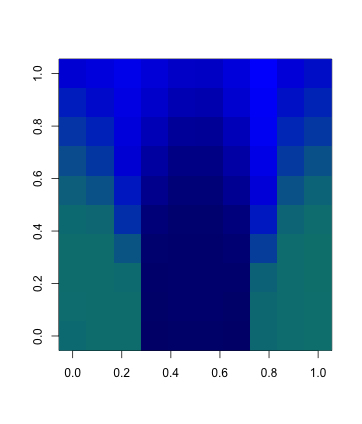}
\includegraphics[width = 0.2\textwidth]{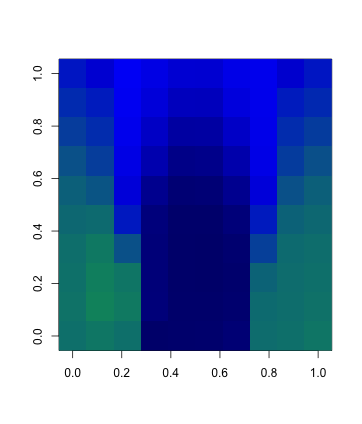}
\includegraphics[width = 0.2\textwidth]{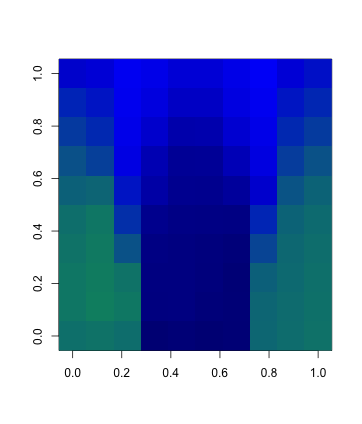}
\includegraphics[width=0.047 \textwidth]{colorbar_2.png}
\end{array}$
\end{center}
\caption{\label{FIG: simulation 2d}2D simulation for mixed Cauchy and mixed Slash distribution. The first row: true function $f_0$; The second row to forth row present the  contaminated data $Y^o$, DNN estimations , RDNN estimations. From left to right, the observed data are generated from  Case 1 (i) and (ii), Case 2 (i) and (ii).}
\end{figure}

Under the proposed functional model (\ref{subSEC:2D}), we introduce outlier hyper-surfaces to the generated functional sample by randomly contaminating a subset, $R^o$, of the original sample. The contamination proportion $r$ is chosen to be $0$, $0.1$ and $0.2$. The similar simulation setting has been considered in \cite{Lima:etal:19,Lima:etal:19b}. We consider the following four types of outliers, i.e., two surface outliers and two heavy-tailed distributed outliers. They mimic the types of noised data   encountered in the real dataset in Section \ref{SEC:realdata}.
 \begin{description}
 \item [Case 1:] {\it Stripe outliers} To simulate outliers on a stripe in 2D regions, the contamination occurs on a line segment $a_0\times \mathcal{I}$, that is,
 \begin{equation*}
     Y_{ij^\ast}^o = Y_{ij^\ast} + \epsilon_{ij^\ast}^o,\;i\in R^o, \;\;\;
     j^\ast_1/N_2=a_0, \;\;\;j^\ast_2/N_2\in  \mathcal{I},
 \end{equation*}
 where $\epsilon_{ij^\ast}^o \sim U\left( 10, 20\right)$. In this simulation,    $a_0=0.2$, and we choose (i) $ \mathcal{I}=\cup_{k=1}^5\left[ \frac{2k-2}{10}, \frac{2k-1}{10}\right)$, and (ii) $ \mathcal{I} = \left[ 0, 1\right]$.

 \item [Case 2:] {\it Square outliers} To simulate outliers on a consecutive 2D region, the contamination occurs on a square $\left[a_0, a_1 \right]^2$, that is,
 \begin{equation*}
     Y_{ij^\ast}^o = Y_{ij^\ast} + \epsilon_{ij^\ast}^o,\;i\in R^o, \;\;\;
     \left(j^\ast_1/N_2, j^\ast_2/N_2\right)\in \left[a_0, a_1 \right]^2
 \end{equation*}
 where $\epsilon_{ij^\ast}^o \sim U\left( 10, 20\right)$.
 In the simulation, we choose (i) $\left[a_0, a_1 \right]^2 = \left[0.1, 0.3 \right]^2$, and (ii) $ \left[a_0, a_1 \right]^2 = \left[0.1, 0.5 \right]^2$.

 \item [Case 3:]{\it Mixture Normal–Cauchy}  To simulate outliers with heavy-tailed distribution,  the distribution of $\epsilon_{ij^\ast}^o$'s follow  a mixture of a normal distribution
$N(0, 1)$ and a Cauchy distribution with location $0$ and scale $0.5$. The mixing weights for Cauchy distribution are (i) $30\%$, and  (ii) $50\%$.

 \item [Case 4:] {\it Mixture Normal–Slash} Similar to previous case, but using a mixture of a normal
distribution $N(0, 1)$ and a Slash distribution with location $0$ and scale $0.5$. The mixing weights for Slash distribution are  (i) $30\%$, and  (ii) $50\%$.
 \end{description}
 We consider sample size $n= 50, 100, 200$. For each image, let $N_2  = 10$, implicating the number of observational points (pixels) is set to be $N=N_2^2 = 100$. The network architecture is determined in a data driven way as suggested in Section \ref{SEC:architecture}, and we use Huber's loss function with tuning parameter $1$ for RDNN estimator in (\ref{EQ:fhat}). The results of each setting are based on $100$ Monte Carlo simulations.
  Figures \ref{FIG: simulation 2d} presents heat maps of a typical set of  the true mean function and abnormal observations, along with the estimations of RDNN and DNN estimators.
  From Table \ref{TAB:Sim2d_uncontamination}, we can see that when training the clean data, DNN method has comparable $L_2$ risks with RDNN estimators. These risks decrease as the sample size $n$ increases. However, when contamination is involved, Table \ref{TAB:Sim2d_contamination} shows that the  risks of DNN estimators elevated drastically, while RDNN ones keep consistent results. In addition, although increasing either contamination rate $r$ or contamination areas on a stripe raises the risks, we can see that RDNN estimators perform steady and remains relatively small $L_2$ risks even given $20\%$ data contain anomalies. From Table \ref{TAB:Sim2d_contamination}, we can also see that when contamination occurs in a square region, the same trend is revealed, as previous discussion. It is worth mentioning that when $r=0.2$, for the contaminated region $\left[0.1, 0.5\right]^2$, DNN estimators has extremely large  risks, which are more than $10$ times of ones of RDNN. Similar findings can be concluded from Table \ref{TAB:Sim2d_nongaussin}, where the random errors following non-Gaussian heavy-tail distributions.
The RDNN estimator best mitigates the effect of this contamination relative to its competitors. Overall, the present simulation experiments suggest that
RDNN perform well in clean data and safeguard against
outlying observations either in the form of outlying surfaces or heavy-tailed measurement errors.

\begin{figure}
\begin{center}
\hspace{-9.2cm}  $f_0$ \hspace{0.25cm}
$\begin{array}{l}
\includegraphics[width = 0.2\textwidth]{b_part_true.png}
\includegraphics[width=0.047 \textwidth]{colorbar_2.png}
\end{array}$\\
outliers
$\begin{array}{l}
\includegraphics[width = 0.2\textwidth]{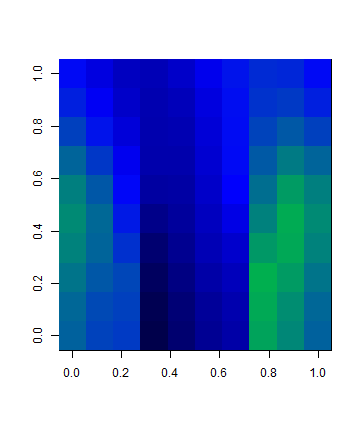}
\includegraphics[width = 0.2\textwidth]{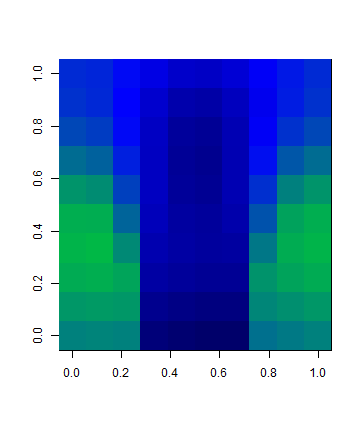}
\includegraphics[width = 0.2\textwidth]{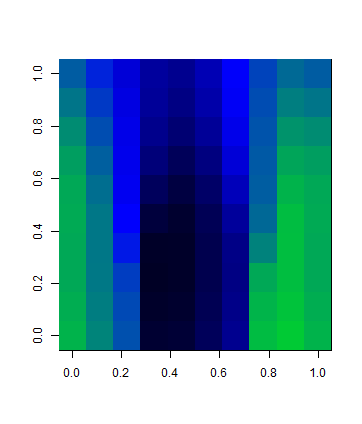}
\includegraphics[width = 0.2\textwidth]{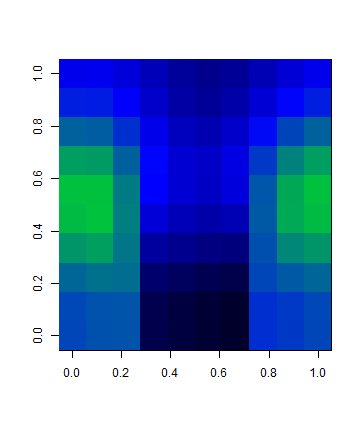}
\includegraphics[width=0.047 \textwidth]{colorbar_2.png}
\end{array}$\\
DNN \;\;
$\begin{array}{l}
\includegraphics[width = 0.2\textwidth]{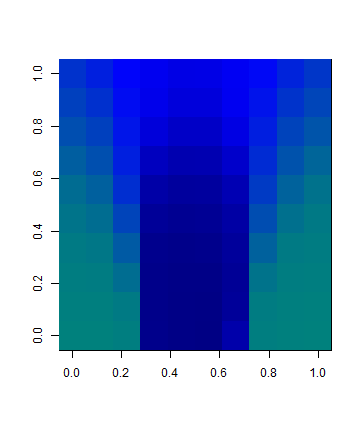}
\includegraphics[width = 0.2\textwidth]{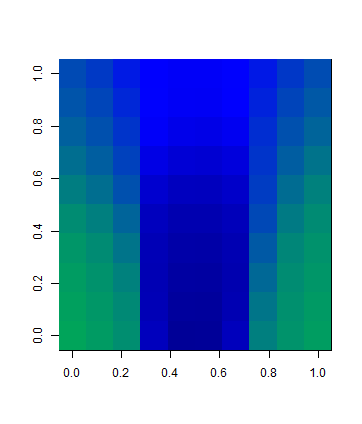}
\includegraphics[width = 0.2\textwidth]{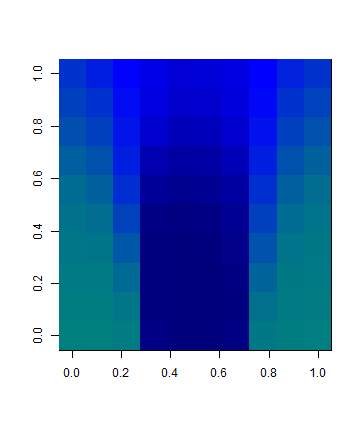}
\includegraphics[width = 0.2\textwidth]{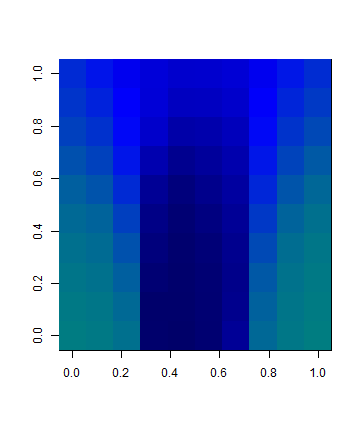}
\includegraphics[width=0.047 \textwidth]{colorbar_2.png}
\end{array}$\\
RDNN
$\begin{array}{l}
\includegraphics[width = 0.2\textwidth]{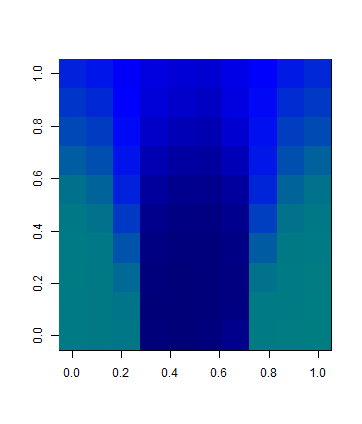}
\includegraphics[width = 0.2\textwidth]{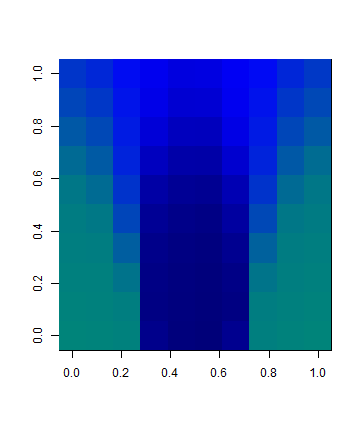}
\includegraphics[width = 0.2\textwidth]{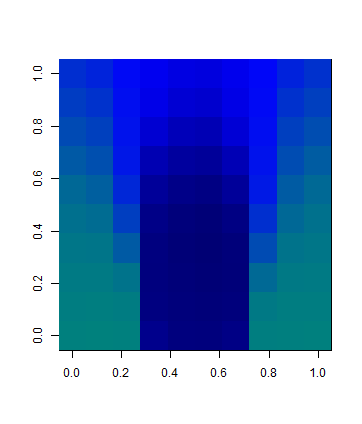}
\includegraphics[width = 0.2\textwidth]{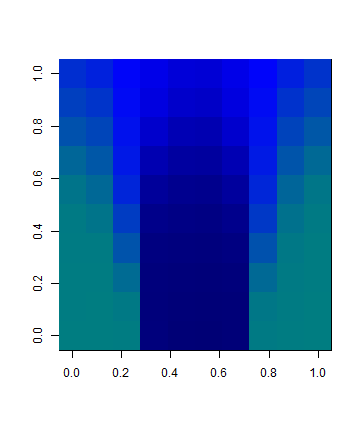}
\includegraphics[width=0.047 \textwidth]{colorbar_2.png}
\end{array}$
\end{center}
\caption{\label{FIG: simulation 2d mixture}2D simulation for mixed Cauchy and mixed Slash distribution. The first row: true function $f_0$; The second row to forth row present the  contaminated data $Y^o$, DNN estimations , RDNN estimations. From left to right, the observed data are generated from  Case 3 (i) and (ii), Case 4 (i) and (ii).}
\end{figure}

\begin{table}[tbh] \centering
  \caption{Empirical $L_2$ risk of 2D uncontaminated data with standard errors in brackets.}
  \label{TAB:Sim2d_uncontamination}
\begin{tabular}{@{\extracolsep{0.1pt}} ccc}
\hline
\hline

$n$ & \multicolumn{1}{c}{RDNN} & \multicolumn{1}{c}{DNN}\\
\hline
50& 0.114 (0.040) & 0.125 (0.049)\\
100& 0.059 (0.029)& 0.055 (0.028)\\
200& 0.034 (0.016)& 0.031 (0.017) \\
\hline
\hline
\end{tabular}
\end{table}

\begin{table}[tbh] \centering
  \caption{Empirical $L_2$ risk of 2D contaminated data in Cases 1 and 2 with standard errors in brackets.}
  \label{TAB:Sim2d_contamination}
\begin{tabular}{@{\extracolsep{0.1pt}} ccccccccc}
\hline
\hline
\multicolumn{2}{c}{contaminated regions}   & $n$&\multicolumn{2}{c}{$r=0.1$} && \multicolumn{2}{c}{$r=0.2$}\\
\cline{4-5} \cline{7-8}
&&   & \multicolumn{1}{c}{RDNN} & \multicolumn{1}{c}{DNN}& &\multicolumn{1}{c}{RDNN} & \multicolumn{1}{c}{DNN}\\
\hline
\multirow{6}{*}{stripe}  & &50 &0.115 (0.048) & 0.179 (0.078)  && 0.128 (0.055)  & 0.329 (0.095)\\
&$\cup_{k=1}^5\left[ \frac{2k-2}{10}, \frac{2k-1}{10}\right)$ &100 & 0.055 (0.023)  & 0.102 (0.033) && 0.065 (0.033) & 0.252 (0.055) \\
&&200 & 0.032 (0.015) & 0.081 (0.024) && 0.041 (0.018) & 0.257 (0.043) \\
\cline {2-8}
&&50 & 0.137 (0.066) & 0.311 (0.081) && 0.151 (0.051)  & 0.864 (0.164) \\
&$\left[ 0, 1\right]$&100 & 0.064 (0.027)  & 0.240 (0.055) && 0.088 (0.029)  & 0.842 (0.112)  \\
&&200 & 0.036 (0.015)  & 0.226 (0.032) && 0.064 (0.023)& 0.848 (0.084)  \\
\cline {1-8}
\multirow{6}{*}{square} &&50 &0.118 (0.048)& 0.260 (0.093) && 0.154 (0.071) & 0.664 (0.170) \\
&$\left[0.1, 0.3 \right]^2$&100 & 0.065 (0.033) & 0.195 (0.059) && 0.069 (0.025) & 0.754 (0.107) \\
&&200 & 0.038 (0.019) & 0.195 (0.042) && 0.054 (0.022) & 0.752 (0.091) \\
\cline {2-8}
&&50 & 0.151 (0.060) & 0.657 (0.159) && 0.234 (0.080)  & 2.014 (0.297) \\
&$\left[0.1, 0.5 \right]^2$&100 & 0.078 (0.042) & 0.533 (0.111) && 0.134 (0.063) & 2.070 (0.248) \\
&&200 & 0.046 (0.023) & 0.550 (0.091) && 0.108 (0.042) & 2.172 (0.191) \\
\hline
\hline
\end{tabular}
\end{table}

\begin{table}[tbh] \centering
  \caption{Empirical $L_2$ risk of 2D contaminated data in Cases 3 and 4 with standard errors in brackets.}
  \label{TAB:Sim2d_nongaussin}
\begin{tabular}{@{\extracolsep{0.1pt}} ccccccc}
\hline
\hline
\multicolumn{1}{c}{error types}   & $n$&\multicolumn{2}{c}{mixing weight$=30\%$} && \multicolumn{2}{c}{mixing weight$=50\%$}\\
\cline{3-4} \cline{6-7}
&   & \multicolumn{1}{c}{RDNN} & \multicolumn{1}{c}{DNN}& &\multicolumn{1}{c}{RDNN} & \multicolumn{1}{c}{DNN}\\
\hline
 &50 & 0.186 (0.069)  & 0.665 (0.959) &&  0.191 (0.069) & 1.343 (3.193)  \\
Cauchy &100 &  0.097 (0.044) &  0.289 (0.265) &&   0.104 (0.586)  & 0.586 (0.799)\\
&200 &  0.051 (0.029) & 0.140 (0.175) &&  0.053 (0.024) & 0.104 (0.066)\\
\cline {1-7}
&50 & 0.142 (0.065) & 0.456 (0.686) && 0.136 (0.071)  & 0.949 (2.022) \\
Slash &100 & 0.074 (0.033)  & 0.419 (0.948) && 0.071 (0.033)  & 0.822 (1.533)  \\
&200 & 0.054 (0.027)  & 0.304 (0.617) && 0.055 (0.029)& 0.544 (1.004)  \\
\hline
\hline
\end{tabular}
\end{table}

 \subsection{3D simulation}\label{subSEC:3D}
For 3D simulation, the functional data are generated from the model (\ref{EQ:sim}).  The true mean function is
$f_0(\mathbf{x}_j)=f_0(x_{1j},x_{2j}, x_{3j})=\exp\left(\frac{1}{3}x_{1j} + \frac{1}{3}x_{2j}+\sqrt{x_{3j}+0.1}\right)$, where $\mathbf{x}_j=(x_{1j}, x_{2j}, x_{3j})= \left(j_1/N_3, j_2/ N_3, j_3/N_3 \right)$, $1 \leq j_1, j_2,j_3 \leq N_3$,   are equally spaced grid points in $\left[ 0, 1\right] ^3$ and $N=N_3^3=5^3$. Generate $\eta(\cdot)$  from a Guassian process, with zero mean and covariance function $G_0(\mathbf{x}_j,\mathbf{x}_{j'})=\sum_{k=1}^3\cos(2\pi( {x}_{kj}- {x}_{kj'}))$, $j,j'=1,\ldots,N$, and the measurement errors
$e_{ij}$'s are i.i.d. random variables generated from standard normal distribution.
To contaminate the clean data, we apply the similar settings in Section \ref{subSEC:2D}.
 \begin{description}
 \item [Case 5] To simulate outliers on a consecutive 3D region, the contamination occurs on a square $\left[a_0, a_1 \right]^3$, that is,
 \begin{equation*}
     Y_{ij^\ast}^o = Y_{ij^\ast} + \epsilon_{ij^\ast}^o,\;i\in R^o, \;\;\;
     \left(j^\ast_1/N_3, j^\ast_2/N_3, j^\ast_3/N_3\right)\in \left[a_0, a_1 \right]^3
 \end{equation*}
 where $\epsilon_{ij^\ast}^o \sim U\left( 10, 20\right)$.
 In the simulation, we choose $\left[a_0, a_1 \right]^3 = \left[0.10, 0.20 \right]^3$ and $  \left[0.10, 0.30 \right]^3$ for different contamination proportions.

 \item [Case 6] {\it Mixture Normal–Cauchy}  Similar to case 3,   the distribution of $\epsilon_{ij^\ast}^o$'s follow  a mixture of a normal distribution $N(0, 1)$ and a Cauchy distribution with location $0$ and scale $0.5$. The mixing weights for Cauchy distribution are  (i) $30\%$, and  (ii) $50\%$.

\item [Case 7:] {\it Mixture Normal–Slash} Similar to case 4,  the distribution of $\epsilon_{ij^\ast}^o$'s follow  a mixture of a normal distribution $N(0, 1)$ and a Slash distribution with location $0$ and scale $0.5$. The mixing weights for Slash distribution are  (i) $30\%$, and  (ii) $50\%$.
 \end{description}


 The results of each setting are based on $100$ Monte Carlo simulations for sample sizes are $50,100$, and $ 200$.
  For reference, Table \ref{TAB:Sim3d_uncontamination_clean} shows the average of empirical $L_2$ risks for clean data.  We find that when data are clean, both of RDNN and DNN provide comparable estimations results, and the empirical  risk decreases as the sample size increases.
 Tables \ref{TAB:Sim3d_contamination_cub} and   \ref{TAB:Sim3d_mix} report the average of empirical $L_2$ risks for cases 6 and 7.  As expected, non-robust DNN estimator has explosive  risks, which are around three times of those for uncontaminated data. Similar to the 2D cases,   either enlarging the contaminated region or the contamination proportion increases    risk with DNN estimators. The precision of the RDNN estimator is kept at the same level as all   outlier types and the clean dataset. This provides strong evidence that the proposed RDNN estimator is less sensitive to the presence of outliers, maintaining precision. In the worst case, the  risks of RDNN estimator have increased no more than four times  compared with the clean data scenarios, however, the non-robust ones have increased around $20$ times.

\begin{table}[tbh] \centering
  \caption{Empirical $L_2$ risk of 3D uncontaminated data with standard errors in brackets.}
  \label{TAB:Sim3d_uncontamination_clean}
\begin{tabular}{@{\extracolsep{0.1pt}} ccc}
\hline
\hline

$n$ & \multicolumn{1}{c}{RDNN} & \multicolumn{1}{c}{DNN}\\
\hline
50&0.103 (0.050) & 0.090 (0.045)\\
100&0.055 (0.033) & 0.047 (0.023)\\
200& 0.027 (0.013) & 0.026 (0.018) \\
\hline
\hline
\end{tabular}
\end{table}

\begin{table}[tbh] \centering
  \caption{Empirical $L_2$ risk of 3D contaminated data for cases 5  with standard errors in brackets.}
  \label{TAB:Sim3d_contamination_cub}
\begin{tabular}{@{\extracolsep{0.1pt}} cccccccc}
\hline
\hline
Contaminated & $n$ & \multicolumn{2}{c}{$r=0.1$} && \multicolumn{2}{c}{$r=0.2$}\\
\cline{3-4} \cline{6-7}
regions&  & \multicolumn{1}{c}{RDNN} & \multicolumn{1}{c}{DNN}& &\multicolumn{1}{c}{RDNN} & \multicolumn{1}{c}{DNN}\\
\hline
&50 & 0.111 (0.049) & 0.204 (0.066) && 0.119 (0.052) & 0.515 (0.107) \\
$\left[0.10, 0.20 \right]^3$&100 & 0.056 (0.028) & 0.155 (0.041) && 0.078 (0.033) & 0.539 (0.067) \\
&200 & 0.033 (0.018) & 0.148 (0.029) && 0.049 (0.017) & 0.571 (0.058) \\
\hline
&50 & 0.118 (0.060) & 0.463 (0.104)  && 0.173 (0.055) & 1.598 (0.212) \\
$\left[0.10, 0.30 \right]^3$&100 & 0.066 (0.032) & 0.472 (0.092) && 0.135 (0.052) & 1.925 (0.160) \\
&200 & 0.042 (0.017) & 0.478 (0.077) && 0.103 (0.033) & 1.942 (0.156) \\
\hline
\hline
\end{tabular}
\end{table}

\begin{table}[tbh] \centering
  \caption{Empirical $L_2$ risk of 3D contaminated data  for cases 6 and 7 with standard errors in brackets.}
  \label{TAB:Sim3d_mix}
\begin{tabular}{@{\extracolsep{0.1pt}} ccccccc}
\hline
\hline
\multicolumn{1}{c}{error types}   & $n$&\multicolumn{2}{c}{mixing weight$=30\%$} && \multicolumn{2}{c}{mixing weight$=50\%$}\\
\cline{3-4} \cline{6-7}
&   & \multicolumn{1}{c}{RDNN} & \multicolumn{1}{c}{DNN}& &\multicolumn{1}{c}{RDNN} & \multicolumn{1}{c}{DNN}\\
\hline
 &50 & 0.130 (0.072) & 0.526 (1.421)  && 0.134 (0.073)  & 0.804 (2.805)\\
Cauchy &100 & 0.066 (0.035)  & 0.459 (0.953) && 0.062 (0.036) & 0.535 (1.295) \\
&200 & 0.043 (0.023) & 0.163 (0.267) && 0.045 (0.026) & 0.418 (0.907) \\
\cline {1-7}
&50 & 0.128 (0.062) &  0.753 (2.220)  && 0.125 (0.057)  & 0.787 (1.938)\\
Slash &100 & 0.066 (0.042)  & 0.403 (0.887) && 0.068 (0.049) &0.760 (1.458) \\
&200 & 0.049 (0.036) & 0.321 (0.771) && 0.047 (0.030) &  0.587 (1.312) \\
\hline
\hline
\end{tabular}
\end{table}

\section{Real data analysis}\label{SEC:realdata}
The  dataset used in the preparation of this article were obtained from the ADNI database (\url{adni.loni.usc.edu}).
The ADNI is a longitudinal multicenter study designed to develop clinical, imaging, genetic, and biochemical biomarkers for the early detection and tracking of AD.
From this database,   {we collect PET data from $85$  patients in AD group. This  PET dataset has been  spatially normalized and post-processed. These AD patients have three to six times doctor visits and we only select the PET scans obtained in the third visits. Patients' age ranges from $59$ to $88$ and average age is $76.49$.  All scans were reoriented into $79\times 95 \times 69$ voxels, which means each patient has $69$ sliced 2D images with  $79\times 95$ pixels.  For 2D case, it indicates that each subject has $N=7,505=79\times 95$ observed pixels for each selected image slice.   
}

\begin{figure}
\hspace{3cm}\textbf{$20$-th}\hspace{2cm}\textbf{$30$-th} \hspace{2.5cm}\textbf{$40$-th}\hspace{2cm}\textbf{$50$-th}\\
Average
$\begin{array}{l}
\includegraphics[width=0.2 \textwidth]{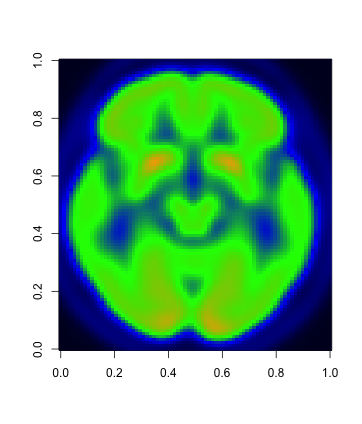}
\includegraphics[width=0.2 \textwidth]{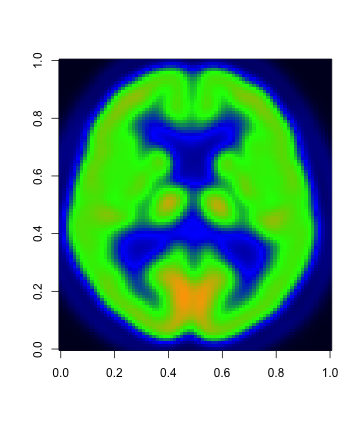}
\includegraphics[width=0.2 \textwidth]{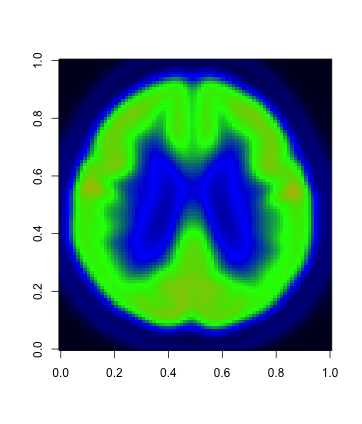}
\includegraphics[width=0.2 \textwidth]{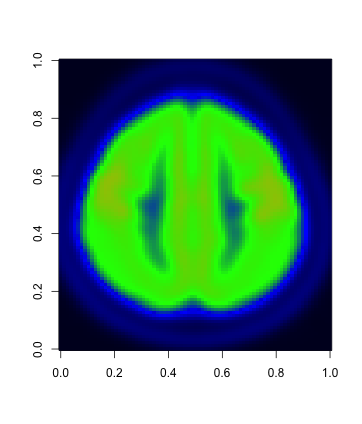}
\includegraphics[width=0.055 \textwidth]{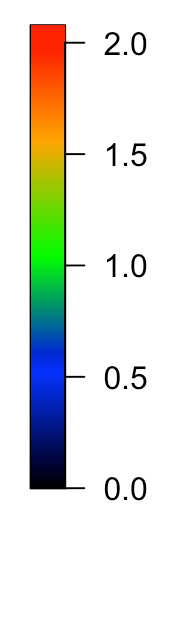}
\end{array}$\\
Outliers\;
$\begin{array}{l}
\includegraphics[width=0.2 \textwidth]{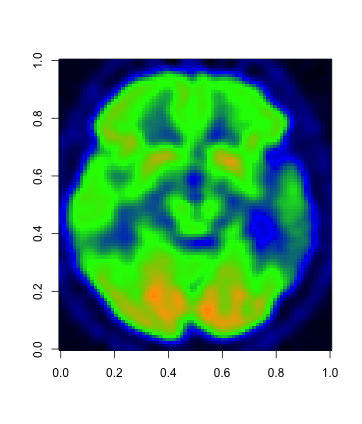}
\includegraphics[width=0.2 \textwidth]{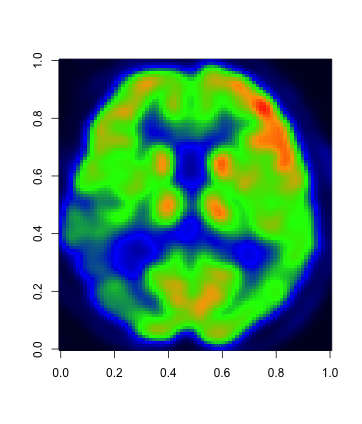}
\includegraphics[width=0.2 \textwidth]{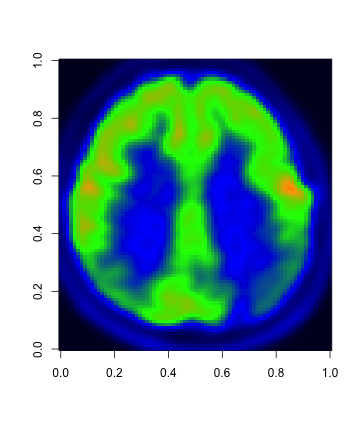}
\includegraphics[width=0.2 \textwidth]{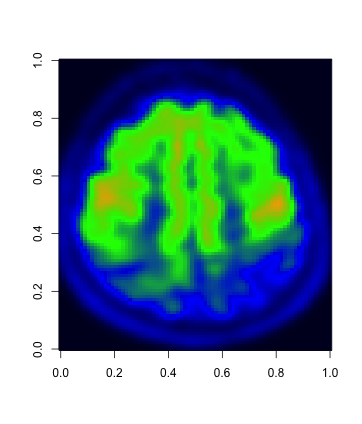}
\includegraphics[width=0.055 \textwidth]{colorbar.png}
\end{array}$\\

\caption{\label{FIG: outlier} The first row are the averaged images for 20-th, 30-th, 40-th and 50-th slices across all patients. The rest are some abnormal data for each slices from some patients. }
\end{figure}

\begin{figure}
\hspace{2cm}\textbf{$20$-th}\hspace{2cm}\textbf{$30$-th} \hspace{2.5cm}\textbf{$40$-th}\hspace{2.5cm}\textbf{$50$-th}\\
$10\%$
$\begin{array}{l}
\includegraphics[width=0.2 \textwidth]{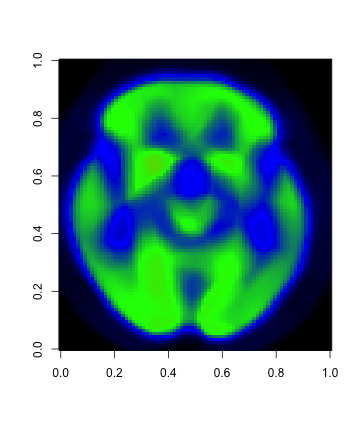}
\includegraphics[width=0.2 \textwidth]{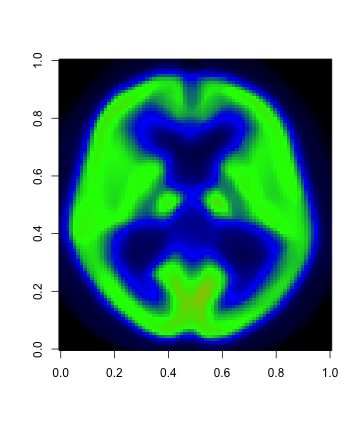}
\includegraphics[width=0.2 \textwidth]{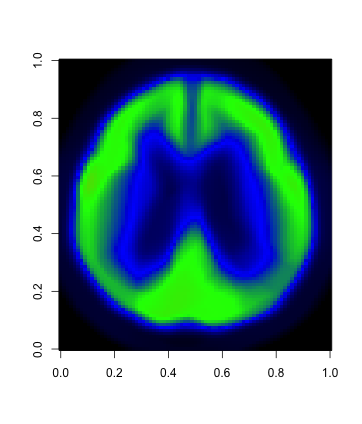}
\includegraphics[width=0.2 \textwidth]{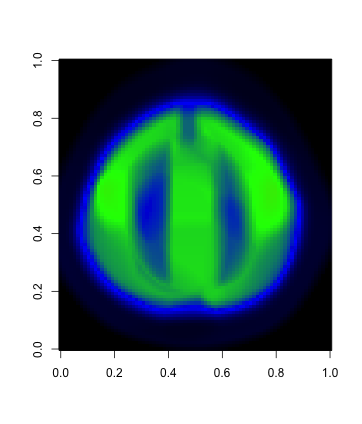}
\includegraphics[width=0.055 \textwidth]{colorbar.png}
\end{array}$\\
$30\%$
$\begin{array}{l}
\includegraphics[width=0.2 \textwidth]{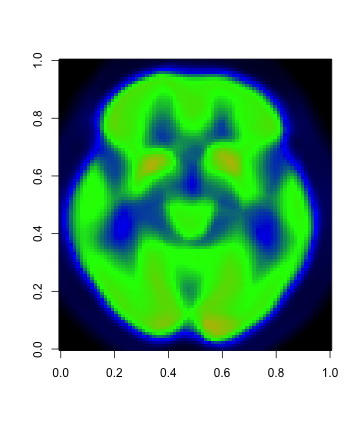}
\includegraphics[width=0.2 \textwidth]{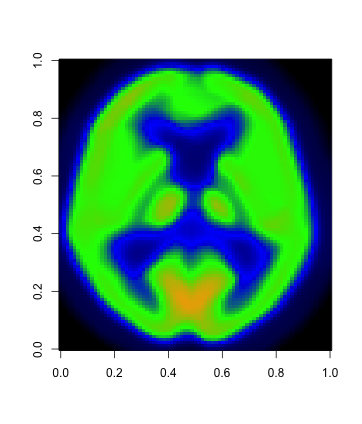}
\includegraphics[width=0.2 \textwidth]{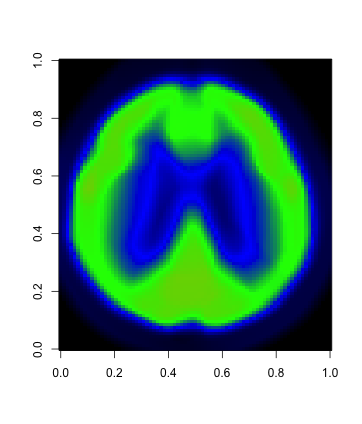}
\includegraphics[width=0.2 \textwidth]{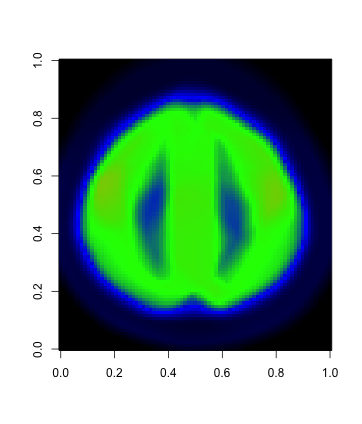}
\includegraphics[width=0.055 \textwidth]{colorbar.png}
\end{array}$\\
$50\%$
$\begin{array}{l}
\includegraphics[width=0.2 \textwidth]{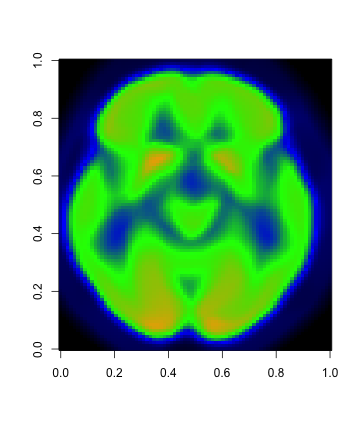}
\includegraphics[width=0.2 \textwidth]{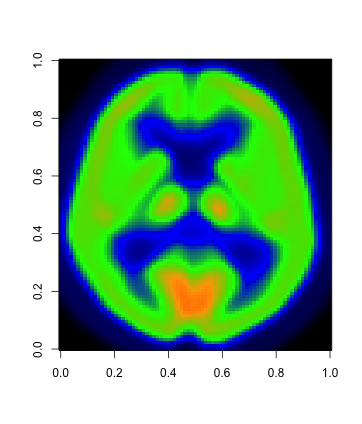}
\includegraphics[width=0.2 \textwidth]{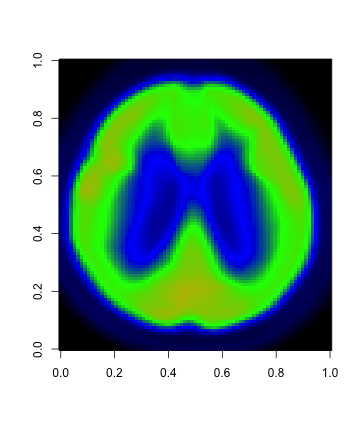}
\includegraphics[width=0.2 \textwidth]{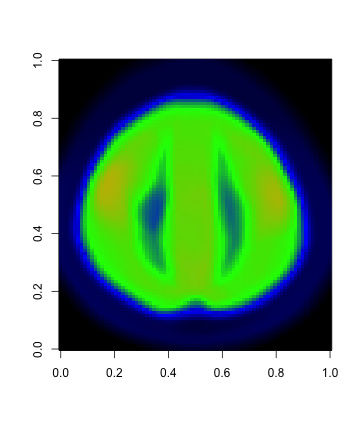}
\includegraphics[width=0.055 \textwidth]{colorbar.png}
\end{array}$\\
$70\%$
$\begin{array}{l}
\includegraphics[width=0.2 \textwidth]{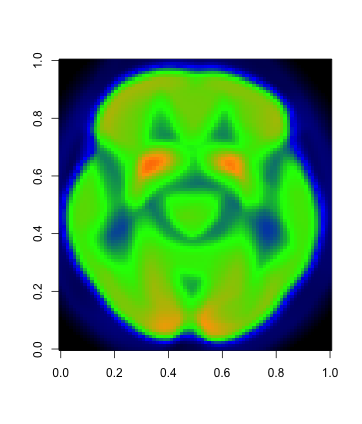}
\includegraphics[width=0.2 \textwidth]{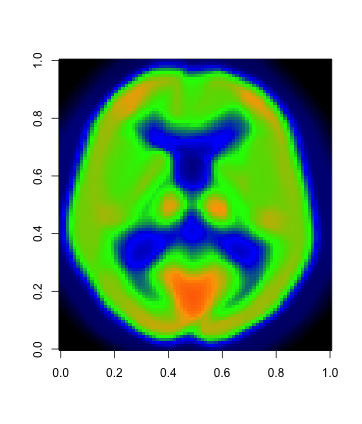}
\includegraphics[width=0.2 \textwidth]{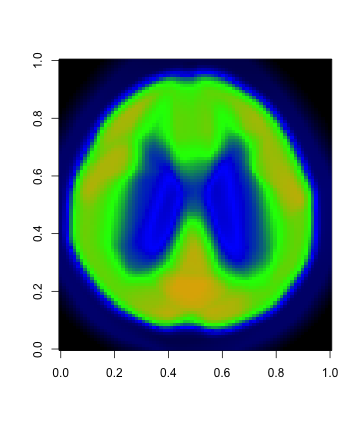}
\includegraphics[width=0.2 \textwidth]{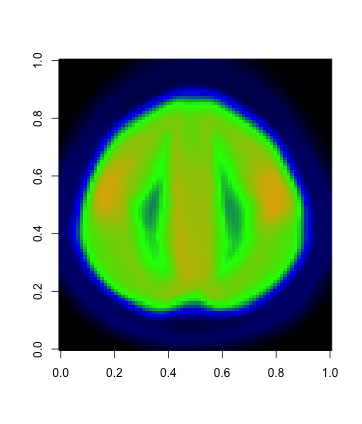}
\includegraphics[width=0.055 \textwidth]{colorbar.png}
\end{array}$\\
$90\%$
$\begin{array}{l}
\includegraphics[width=0.2 \textwidth]{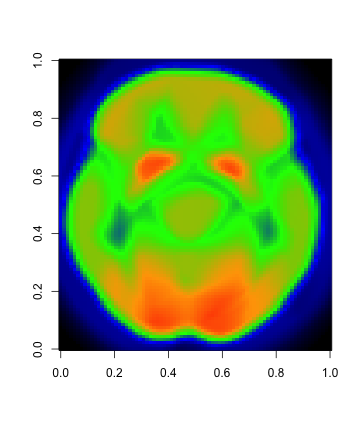}
\includegraphics[width=0.2 \textwidth]{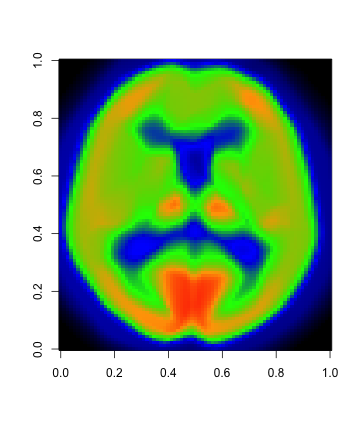}
\includegraphics[width=0.2 \textwidth]{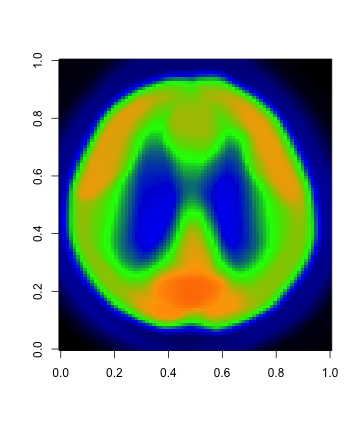}
\includegraphics[width=0.2 \textwidth]{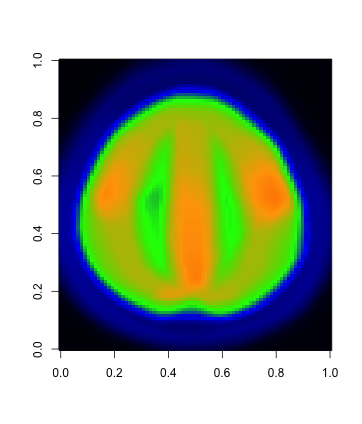}
\includegraphics[width=0.055 \textwidth]{colorbar.png}
\end{array}$
\caption{\label{FIG: 2d_79_95}2D quantile esimators with $79\times 95$ pixels.  From the left to the right: the $20$-th, $30$-th, $40$-th, and $50$-th slices. From the top to the bottom: ($10\%$ , $30\%$ , $50\%$ , $70\%$, $90\%$)-quantiles. }
\end{figure}

\begin{figure}
\hspace{2cm}\textbf{$20$-th}\hspace{2.5cm}\textbf{$30$-th} \hspace{2.5cm}\textbf{$40$-th}\hspace{2.5cm}\textbf{$50$-th}\\
$10\%$
$\begin{array}{l}
\includegraphics[width=0.2 \textwidth]{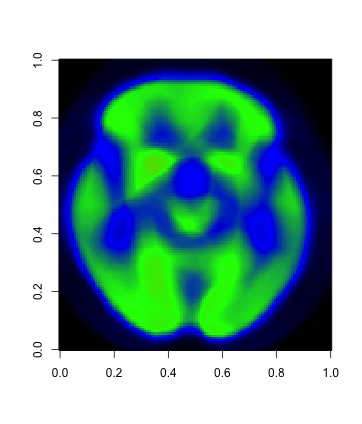}
\includegraphics[width=0.2 \textwidth]{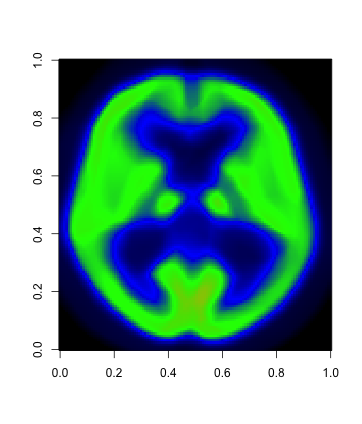}
\includegraphics[width=0.2 \textwidth]{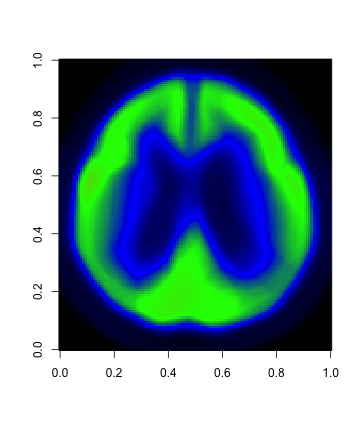}
\includegraphics[width=0.2 \textwidth]{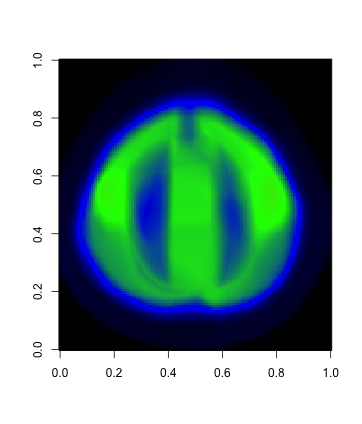}
\includegraphics[width=0.055 \textwidth]{colorbar.png}
\end{array}$\\
$30\%$
$\begin{array}{l}
\includegraphics[width=0.2 \textwidth]{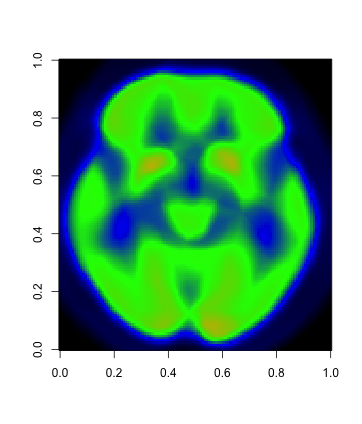}
\includegraphics[width=0.2 \textwidth]{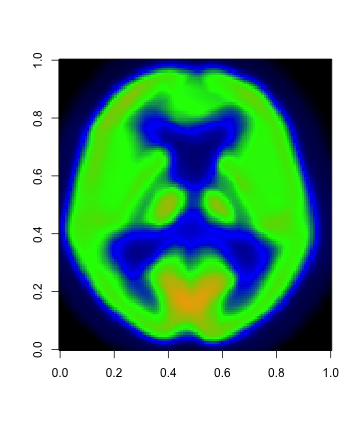}
\includegraphics[width=0.2 \textwidth]{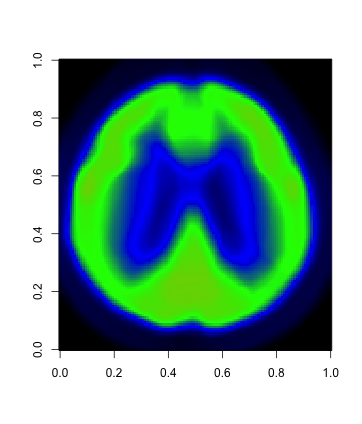}
\includegraphics[width=0.2 \textwidth]{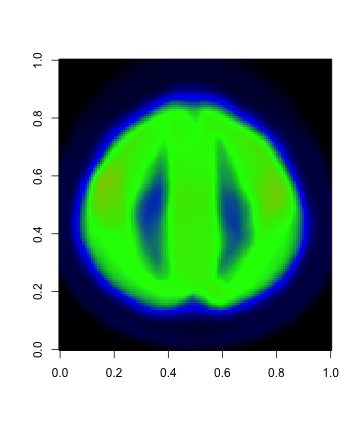}
\includegraphics[width=0.055 \textwidth]{colorbar.png}
\end{array}$\\
$50\%$
$\begin{array}{l}
\includegraphics[width=0.2 \textwidth]{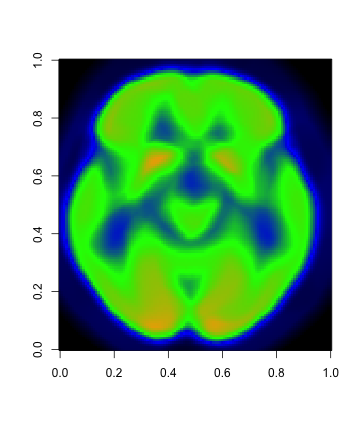}
\includegraphics[width=0.2 \textwidth]{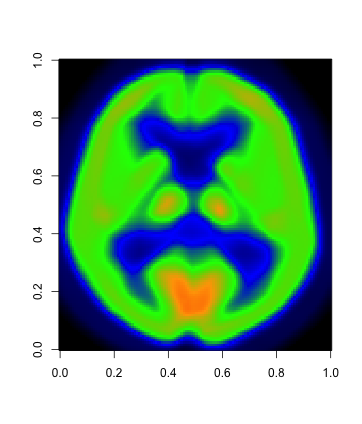}
\includegraphics[width=0.2 \textwidth]{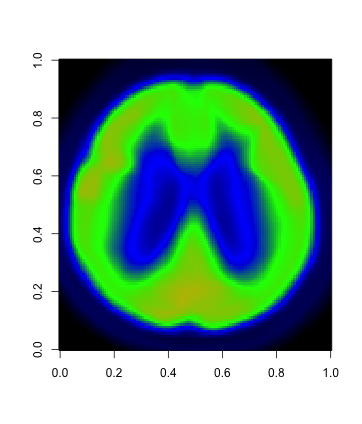}
\includegraphics[width=0.2 \textwidth]{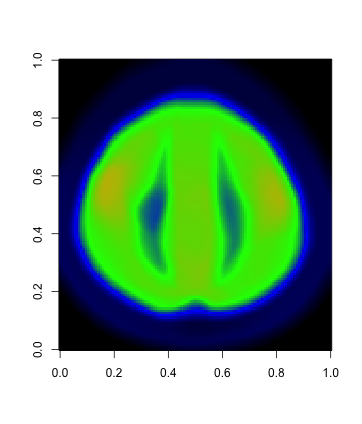}
\includegraphics[width=0.055 \textwidth]{colorbar.png}
\end{array}$\\
$70\%$
$\begin{array}{l}
\includegraphics[width=0.2 \textwidth]{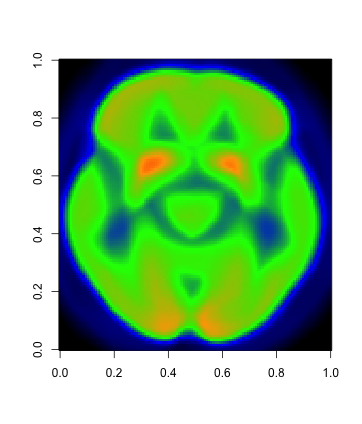}
\includegraphics[width=0.2 \textwidth]{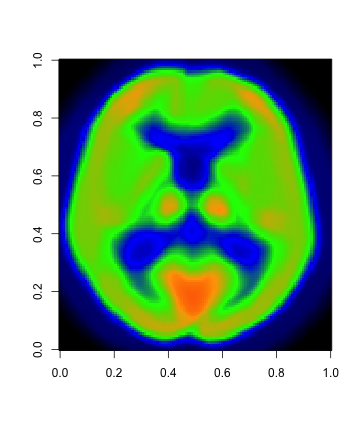}
\includegraphics[width=0.2 \textwidth]{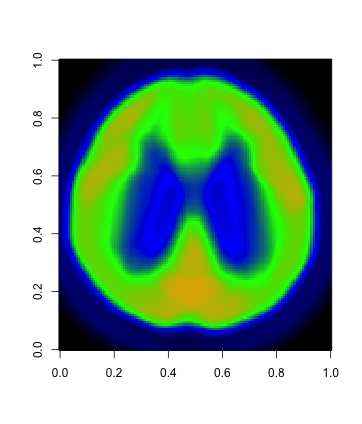}
\includegraphics[width=0.2 \textwidth]{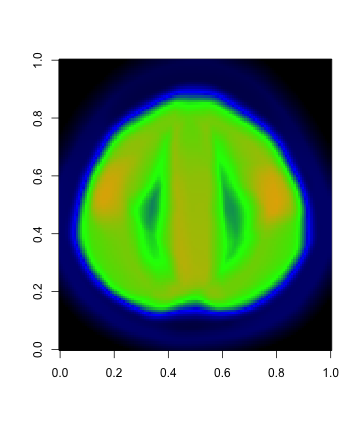}
\includegraphics[width=0.055 \textwidth]{colorbar.png}
\end{array}$\\
$90\%$
$\begin{array}{l}
\includegraphics[width=0.2 \textwidth]{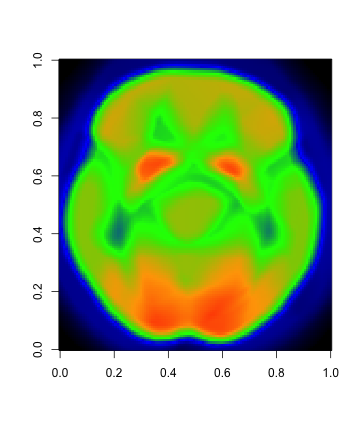}
\includegraphics[width=0.2 \textwidth]{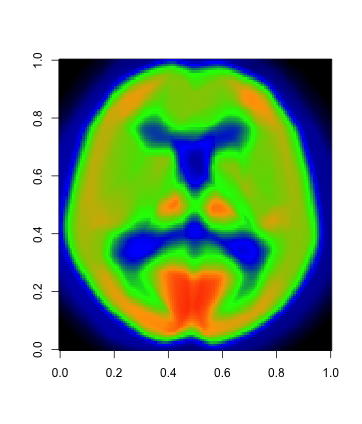}
\includegraphics[width=0.2 \textwidth]{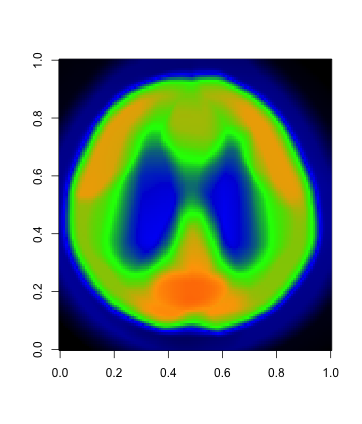}
\includegraphics[width=0.2 \textwidth]{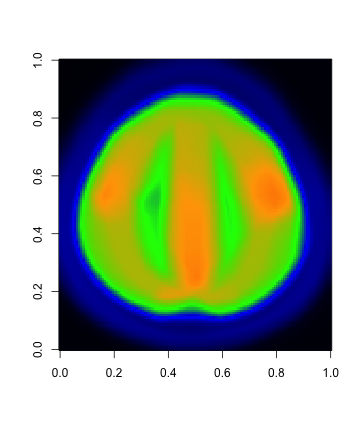}
\includegraphics[width=0.055 \textwidth]{colorbar.png}
\end{array}$
\caption{\label{FIG: 2d_128_128}2D quantile esimators with $128\times 128$ pixels.  From the left to the right: the $20$-th, $30$-th, $40$-th, and $50$-th slices. From the top to the bottom:  ($10\%$ , $30\%$ , $50\%$ , $70\%$, $90\%$)-quantiles. }
\end{figure}

\begin{figure}
\hspace{2cm}\textbf{$20$-th}\hspace{2.5cm}\textbf{$30$-th} \hspace{2.5cm}\textbf{$40$-th}\hspace{2.5cm}\textbf{$50$-th}\\
$10\%$
$\begin{array}{l}
\includegraphics[width=0.2 \textwidth]{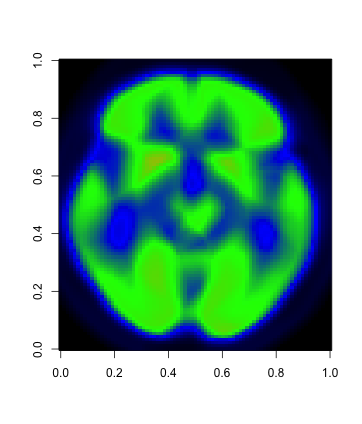}
\includegraphics[width=0.2 \textwidth]{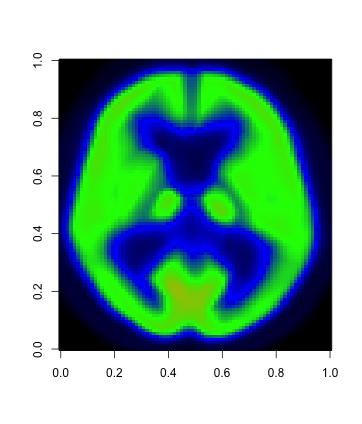}
\includegraphics[width=0.2 \textwidth]{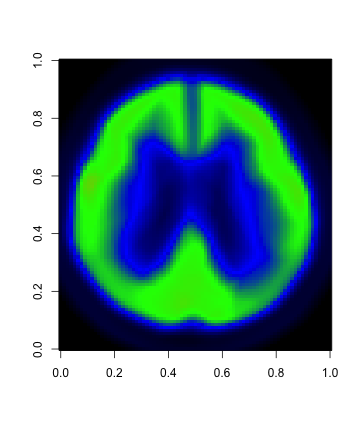}
\includegraphics[width=0.2 \textwidth]{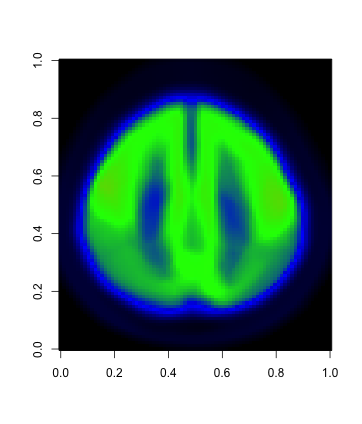}
\includegraphics[width=0.055 \textwidth]{colorbar.png}
\end{array}$\\
$30\%$
$\begin{array}{l}
\includegraphics[width=0.2 \textwidth]{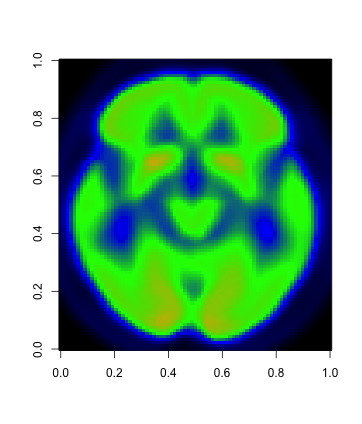}
\includegraphics[width=0.2 \textwidth]{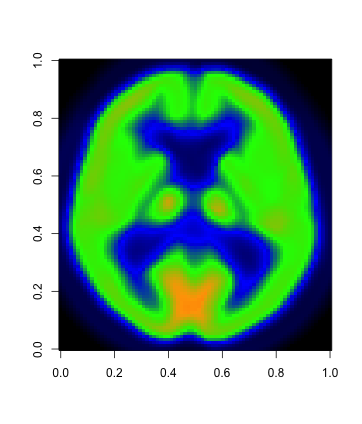}
\includegraphics[width=0.2 \textwidth]{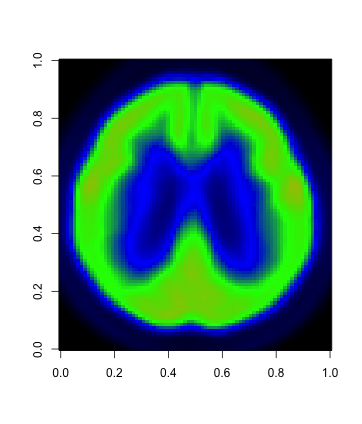}
\includegraphics[width=0.2 \textwidth]{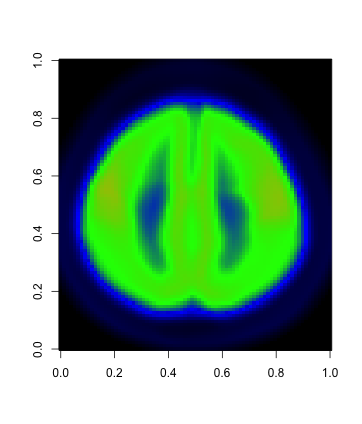}
\includegraphics[width=0.055 \textwidth]{colorbar.png}
\end{array}$\\
$50\%$
$\begin{array}{l}
\includegraphics[width=0.2 \textwidth]{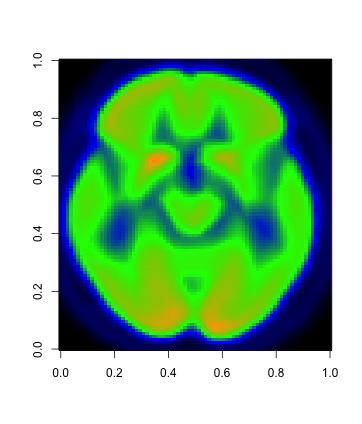}
\includegraphics[width=0.2 \textwidth]{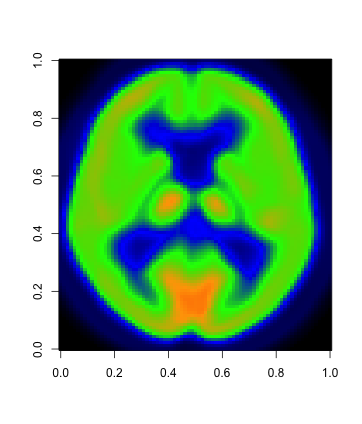}
\includegraphics[width=0.2 \textwidth]{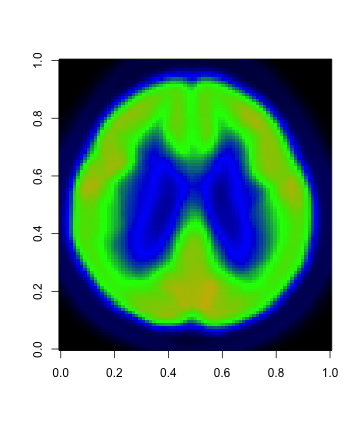}
\includegraphics[width=0.2 \textwidth]{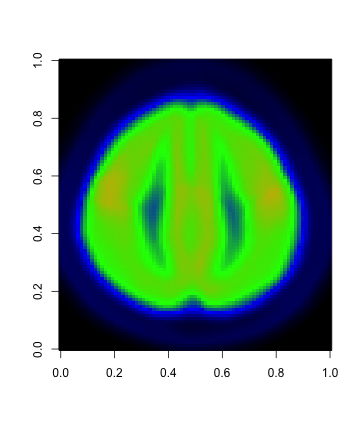}
\includegraphics[width=0.055 \textwidth]{colorbar.png}
\end{array}$\\
$70\%$
$\begin{array}{l}
\includegraphics[width=0.2 \textwidth]{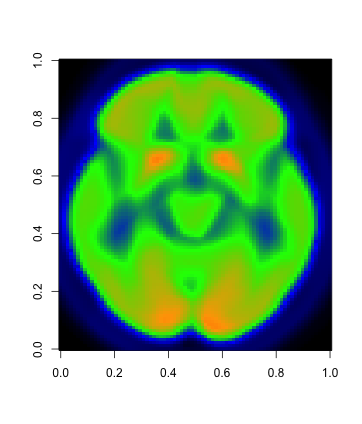}
\includegraphics[width=0.2 \textwidth]{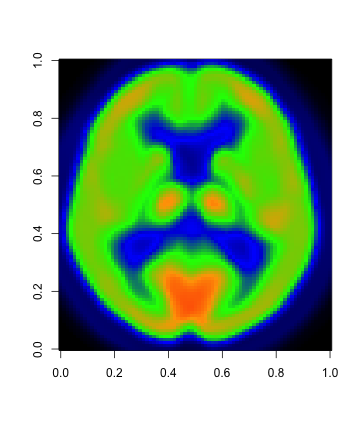}
\includegraphics[width=0.2 \textwidth]{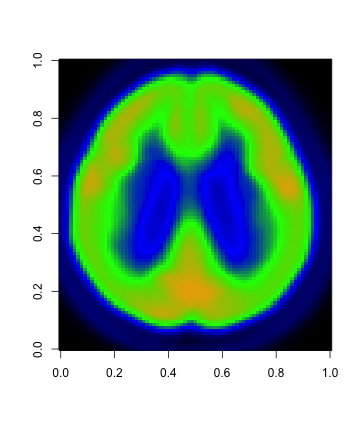}
\includegraphics[width=0.2 \textwidth]{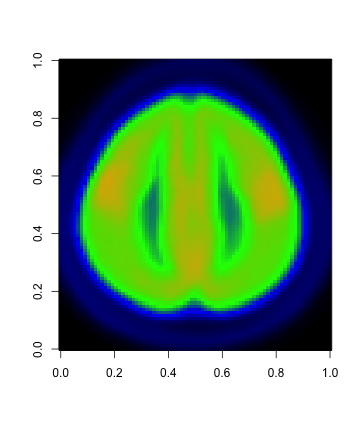}
\includegraphics[width=0.055 \textwidth]{colorbar.png}
\end{array}$\\
$90\%$
$\begin{array}{l}
\includegraphics[width=0.2 \textwidth]{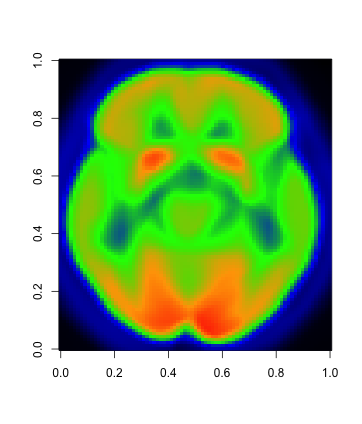}
\includegraphics[width=0.2 \textwidth]{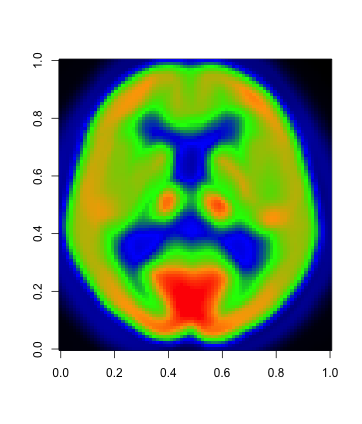}
\includegraphics[width=0.2 \textwidth]{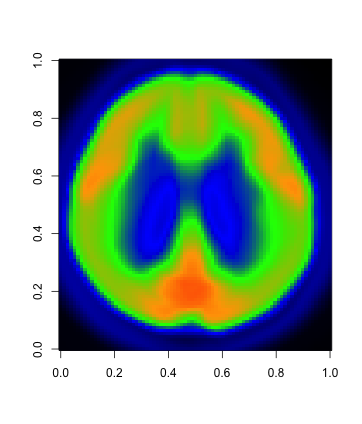}
\includegraphics[width=0.2 \textwidth]{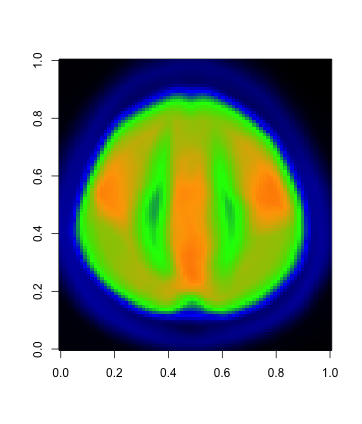}
\includegraphics[width=0.055 \textwidth]{colorbar.png}
\end{array}$
\caption{\label{FIG: 3d_79_95}3D quantile estimators with $79\times 95$ pixels.  From the left to the right: the $20$-th, $30$-th, $40$-th, and $50$-th slices. From the top to the bottom: ($10\%$ , $30\%$ , $50\%$ , $70\%$, $90\%$)-quantiles. }
\end{figure}

\begin{figure}
\hspace{2cm}\textbf{$20$-th}\hspace{2.5cm}\textbf{$30$-th} \hspace{2.5cm}\textbf{$40$-th}\hspace{2.5cm}\textbf{$50$-th}\\
$10\%$
$\begin{array}{l}
\includegraphics[width=0.2 \textwidth]{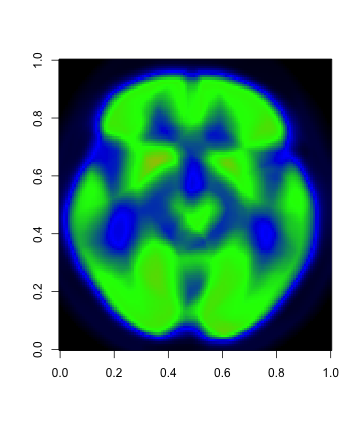}
\includegraphics[width=0.2 \textwidth]{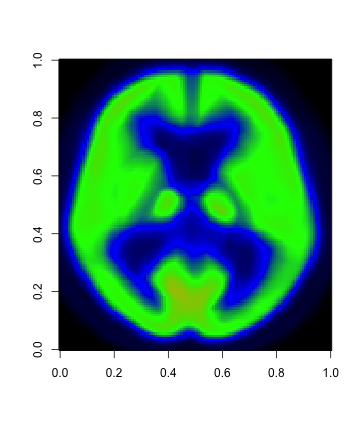}
\includegraphics[width=0.2 \textwidth]{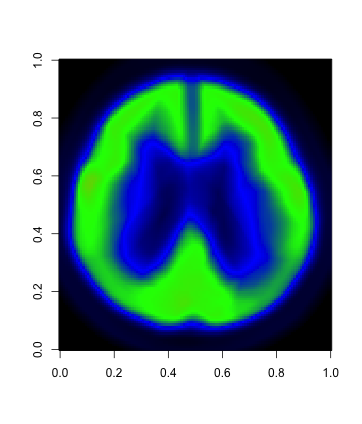}
\includegraphics[width=0.2 \textwidth]{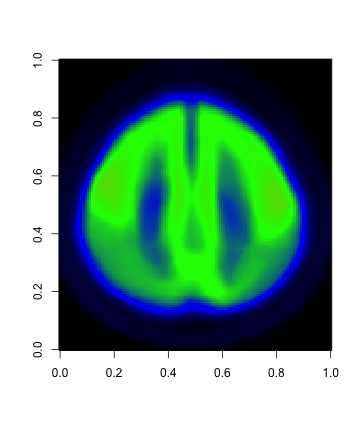}
\includegraphics[width=0.055 \textwidth]{colorbar.png}
\end{array}$\\
$30\%$
$\begin{array}{l}
\includegraphics[width=0.2 \textwidth]{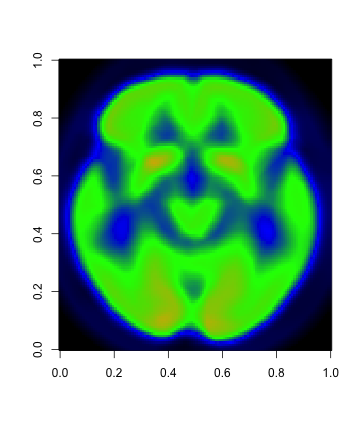}
\includegraphics[width=0.2 \textwidth]{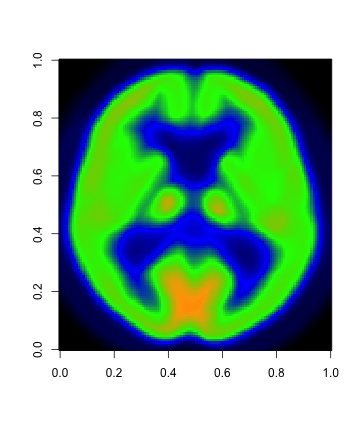}
\includegraphics[width=0.2 \textwidth]{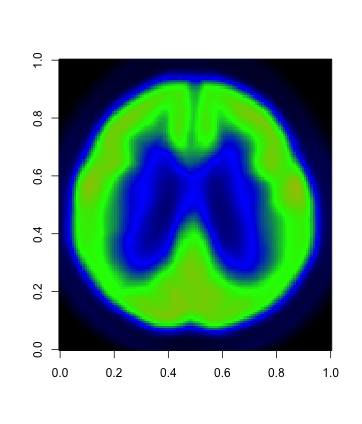}
\includegraphics[width=0.2 \textwidth]{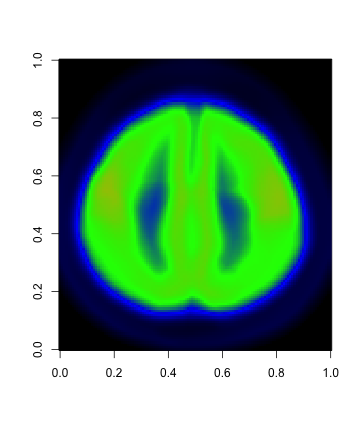}
\includegraphics[width=0.055 \textwidth]{colorbar.png}
\end{array}$\\
$50\%$
$\begin{array}{l}
\includegraphics[width=0.2 \textwidth]{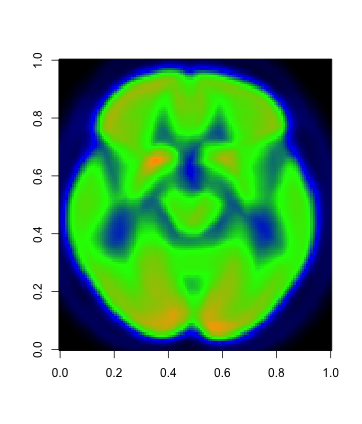}
\includegraphics[width=0.2 \textwidth]{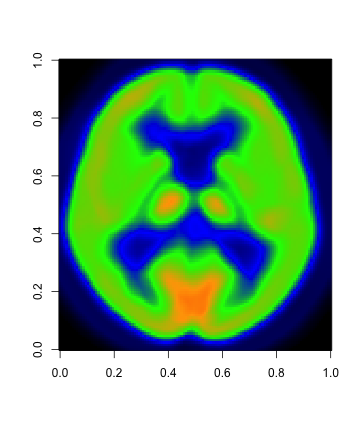}
\includegraphics[width=0.2 \textwidth]{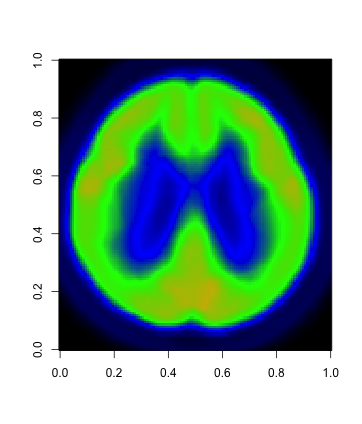}
\includegraphics[width=0.2 \textwidth]{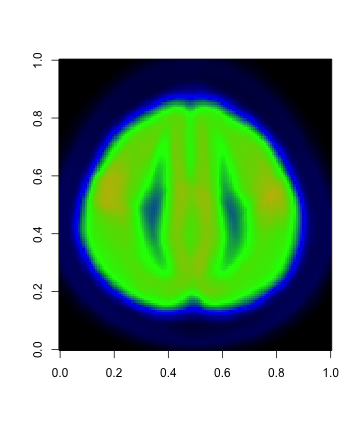}
\includegraphics[width=0.055 \textwidth]{colorbar.png}
\end{array}$\\
$70\%$
$\begin{array}{l}
\includegraphics[width=0.2 \textwidth]{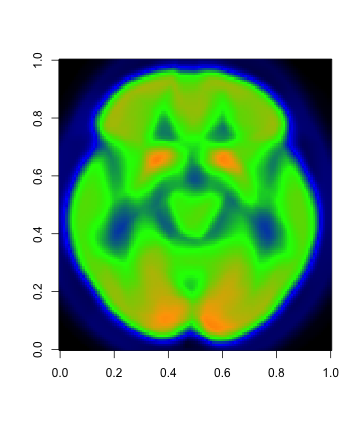}
\includegraphics[width=0.2 \textwidth]{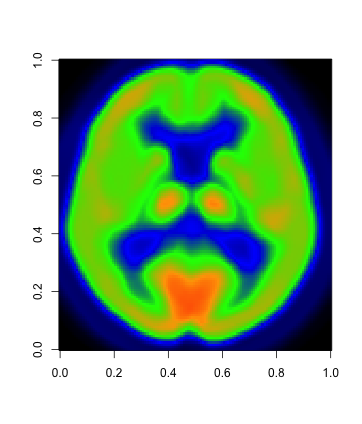}
\includegraphics[width=0.2 \textwidth]{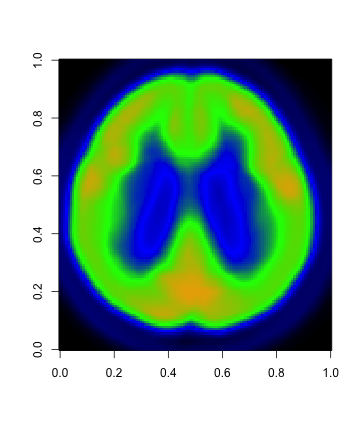}
\includegraphics[width=0.2 \textwidth]{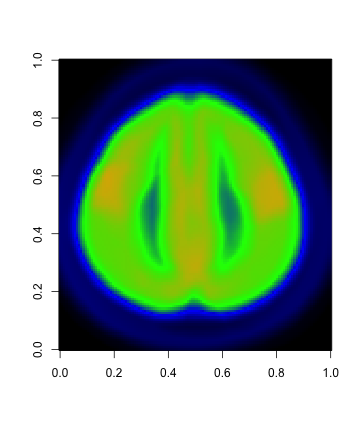}
\includegraphics[width=0.055 \textwidth]{colorbar.png}
\end{array}$\\
$90\%$
$\begin{array}{l}
\includegraphics[width=0.2 \textwidth]{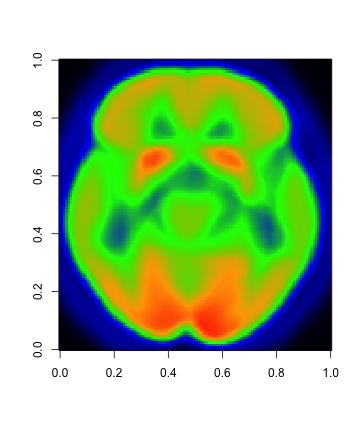}
\includegraphics[width=0.2 \textwidth]{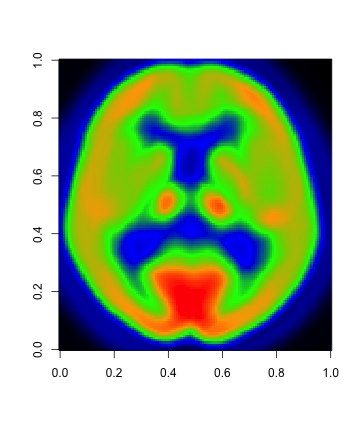}
\includegraphics[width=0.2 \textwidth]{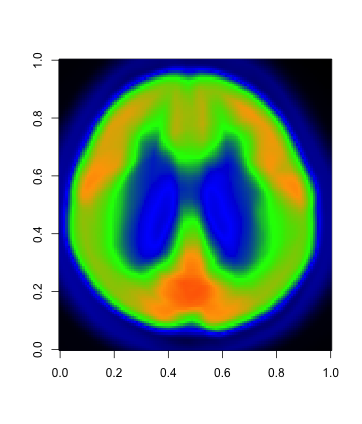}
\includegraphics[width=0.2 \textwidth]{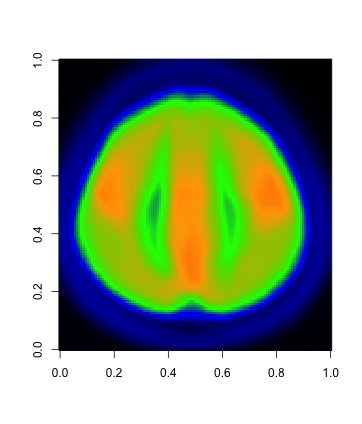}
\includegraphics[width=0.055 \textwidth]{colorbar.png}
\end{array}$
\caption{\label{FIG: 3d_128_128}3D quantile estimators with $128\times 128$ pixels.  From the left to the right: the $20$-th, $30$-th, $40$-th, and $50$-th slices. From the top to the bottom:  ($10\%$ , $30\%$ , $50\%$ , $70\%$, $90\%$)-quantiles. }
\end{figure}

 In this imaging dataset, we observe that there exists a few abnormal  observations, which have different pattern from the majority of data. In Figure \ref{FIG: outlier}, the first row demonstrates the averaged images of the $20$-th, $30$-th, $40$-th, and $50$-th slices across all patients.   In the second row, images are taken from different individuals, where  extreme small values showing in certain regions, which lead to blur boundaries. 
 For the 2D case, we select the $20$-th, $30$-th, $40$-th and $50$-th  slices from $69$ slices for each patient, and apply the proposed RDNN for each slice, respectively, with loss function $\rho_{\tau}(x)=x\left( \tau-\mathbb{I}(x<0)\right)$ with $\tau=0.1,0.3,0.5,0.7,0.9$. The neural network (\ref{EQ:fhat}) is trained through optimizer \texttt{Adam}
with architecture parameters $(L,p,s)$ selected as discussed in Section \ref{SEC:architecture}. We used  $100$ epochs and $128$ as batch size given different data. Based on the images, we obtain the proposed RDNN estimators for each  slice, and also recover the image with the original resolution $79\times 95$ pixels and a higher resolution $128\times 128$. To visualize the estimates,  Figures \ref{FIG: 2d_79_95}  provides the heat maps of the RDNN estimator of different quantiles for all four slices in 2D scenario, Figure \ref{FIG: 2d_128_128} depicts the same estimates but with a  finer resolution ($128\times 128$).  For 3D scenario, we combine all the four slices together, hence, the   3D data totally contains $79\times 95\times 4$ voxels. We first obtain the RDNN estimators with the original resolution and recover them also in a higher resolution $128\times 128\times 4$.  Figures \ref{FIG: 3d_79_95} and \ref{FIG: 3d_128_128} depict the RDNN estimators in the original resolution and higher resolution for each slice and quantile, respectively. The estimated quantiles serve to confirm the suspected multi-modality in this imaging data. According to the heat maps, in $20$-th, $30$-th, and $40$-th slices, higher quantiles significantly differ from lower ones in that there are much larger value presenting in the bottom regions. In particular, for $50$-th slice, higher quantiles can be easily distinguished    from lower ones in terms of overall larger values and wider boundaries.

\section{Discussion}\label{SEC: discussion}
In this work, we resolve the  robust estimation for functional data on multi-dimensional domains via the promising technique from the deep learning.
 By properly choosing network
architecture, our estimator achieves the optimal nonparametric convergence rate in empirical norm.   To the best of our knowledge, the present work is the first work on multi-dimensional functional data robust estimation with   theoretical justification for robust deep learning estimators. Numerical analysis demonstrates that our
approach is useful in recovering the signal for   imaging data given the existing of anomalies.

\section*{Acknowledgements}
Wang's and  Cao's  research was partially supported by  NSF award DMS 1736470.  Cao's  research was also partially supported by  Simons Foundation under Grant \#849413.

Data used in preparation of this article were obtained from the Alzheimers Disease Neuroimaging Initiative (ADNI) database (\url{adni.loni.usc.edu}). As such, the investigators within the ADNI contributed to the design and implementation of
ADNI and/or provided data but did not participate in analysis or writing of this report.
A complete listing of ADNI investigators can be found at: \url{http://adni.loni.usc.edu/wp-content/uploads/how_to_apply/ADNI_Acknowledgement_List.pdf}.


\bibliographystyle{plain} 
\bibliography{Ref}

\end{document}